\pdfoutput=1
\documentclass[%
 aip,
 sd,%
 amsmath,amssymb,
preprint,%
]{revtex4-1}
\usepackage{epsfig}
\usepackage{epstopdf}
\usepackage{graphics}
\usepackage{dcolumn}
\usepackage{bm}
\usepackage{graphicx}
\usepackage{subfigure}

\begin{document}
 
\title{Performance of two-dimensional tidal turbine arrays in free surface flow}

\author{Xianliang Gong}
\author{Ye Li}%
 \email{ye.li@sjtu.edu.cn.}
 \author{Zhiliang Lin}
\author{Qiaohao Hu}
\affiliation{ 
School of Naval Architecture, Ocean \& Civil Engineering, Shanghai Jiao Tong university, Shanghai 200240, China
}%

\begin{abstract}
Encouraged by recent studies on the performance of tidal turbine arrays, we extend the classical momentum actuator disc theory to include the free surface effects and allow the vertical arrangement of turbines. Most existing literatures concern one dimensional arrays with single turbine in the vertical direction, while the arrays in this work are two dimensional (with turbines in both the vertical and lateral directions) and also partially block the channel which width is far larger than height. The vertical mixing of array scale flow is assumed to take place much faster than lateral one. This assumption has been verified by numerical simulations.
Fixing the total turbine area and utilized width, the comparison between two-dimensional and traditional one-dimensional arrays is investigated. The results suggest that the two dimensional  arrangements of smaller turbines are preferred to one dimensional arrays from both the power coefficient and efficiency perspectives. When channel dynamics are considered, the power increase would be partly offset according to the parameters of the channel but the optimal arrangement is unchangeable. Furthermore, we consider how to arrange finite number of turbines in a channel. It is shown that an optimal distribution of turbines in two directions is found. Finally, the scenario of arranging turbines in infinite flow, which is the limiting condition of small blockages, is analysed. A new maximum power coefficient 0.869 occurs when $Fr=0.2$, greatly increasing the peak power compared with existing results.
\end{abstract}

\maketitle

\maketitle

\section{Introduction}
Various types of tidal current turbines are tested and operated around the world. To improve the tidal current turbine's competitiveness in the energy area, it is recommended to rely on tidal turbine arrays as they are more cost-effective than single turbines. Therefore, how to distribute turbines in an array becomes the key point, and a number of important investigations were reported in the past decades. Some people studied tidal arrays by experimental tests \citep[e.g.][]{Stallard2013, Myers2012, Cooke2015} and numerical simulations. \citep[e.g.][]{Churchfield2013, Nishino2013} These detailed approaches provide comprehensive results about tidal arrays, for example the velocity profile of the wake. Another option is theoretical modelling, including the momentum actuator disc theory used in this work. Considering the high computation cost and the difficulty of conducting an experiment with a large number of turbines, theoretical modelling is still the most efficient way to analyse the performance of tidal turbine arrays, especially the limitation of energy extraction.

In recent years, theoretical method has been improved greatly. \citet{Garrett2007} (hereafter GC07) extended the Lanchester-Betz theory by incorporating the blockage effects and introducing the downstream mixing in which the flow is constrained between two rigid surfaces. Importantly, the extra power occurs due to the blockage effects in this model. Neglecting the free surface effects, the application of the GC07 model is somewhat restricted. \citet{Houlsby2008} and \citet{Whelan2009} then developed a model that allows for the deformation of free surface due to the energy extraction. Wake mixing of the free surface model is also considered by \citet{Houlsby2008} to calculate the efficiency (different from power coefficient, here efficiency, also called basin efficiency in some papers, is defined as the fraction of extracted energy to total lost energy in channel) of tidal turbines and the water depth change. 
\begin{figure}
  \centerline{\includegraphics[width=12cm]{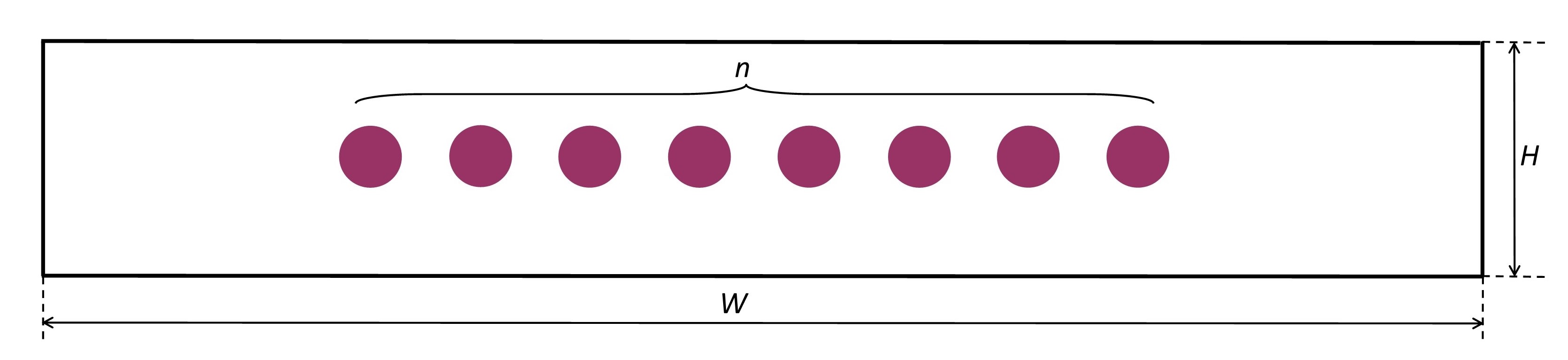}}
  \centerline{(a)}
  \centerline{\includegraphics[width=12cm]{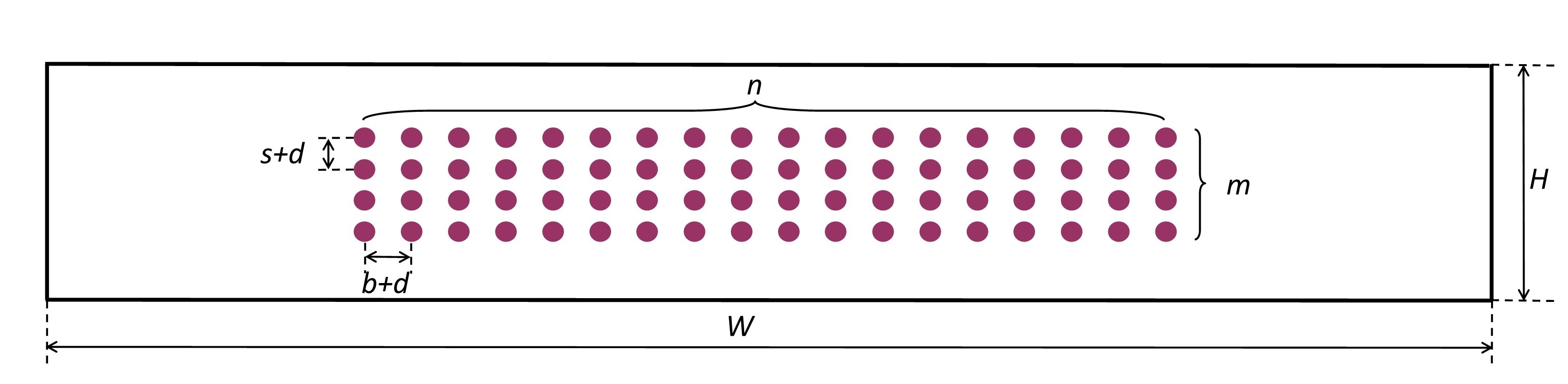}}
  \centerline{(b)}
  \caption{Cross-sectional view of the one-dimensional (which specifically refers to the lateral direction in this work) array with one large turbine in vertical direction (a) and the two-dimensional (which specifically refers to the lateral and vertical direction) array with smaller turbines (b).}
\end{figure}
It should be addressed that the above two single turbine models which can directly be used to analyse the scenario of arranging turbines regularly across the whole width, assume the arrangement of turbines does not change the amount of flux in the 
. In fact, the flow is reduced when the hydrodynamic drag is not negligible. \citet{Garrett2005} (here after GC05) developed a one-dimensional channel dynamical equation to evaluate the interaction between turbines and flows along the channel. \citet{Vennell2010} and \citet{Vennell2011} then combined the GC07 model with the channel dynamical model to figure out the optimal tuning and arrangement of turbines arrayed across the whole width of channel.

As with well known, it is nearly impossible for an array to occupy the whole width of a channel (as above models hypothesized) for reasons like navigations of vessels, environmental requirements and geometrical constraints. More recently, \citet{Nishino2012} (hereafter NW12) derived a two scale rigid lid model (assuming rigid boundaries and neglecting free surface effects) to predicting the performance of the partial tidal array by introducing an idea of scale separation between the flow around each device and the array. In fact, besides the drag caused by turbines, the flow in front of the partial array can also be reduced because of the scale separation. The influence of the array's configuration on performance of turbines has been investigated and the results show that there is an optimal intra-turbine spacing considering power coefficient. 
Including the effects of the array scale flow expansion on local scale dynamics, \citet{Nishino2013} further improved the NW12 model to investigate the performance of whatever long or short fence with arbitrary number of turbines in it. Aiming at the combined effects of two scale flows, channel dynamics are ignored in these two partial array models. 

In summary, the existing literatures investigate arrays with only one turbine in vertical direction, which is the common condition for turbines especially deployed in shallow channel (Fig. 1 (a)). However, turbines are also likely to be placed in much stronger and deeper tidal flows such as Pentland Firth where depth can approach 100 m. In this case, it is difficult to construct, install and maintain a so large turbine to cover the vertical flow area. 
Therefore, it is recommended that several relatively smaller devices be arranged in vertical direction (Fig. 1 (b)) to cover more area (Here we make a definition that this kind of arrays are called as two-dimensional arrays because they are arranged in both the lateral direction and vertical direction compared with one-dimensional arrays which deployed only in the lateral direction.). 
More importantly, turbines in the two-dimensional array show better performances than those in the traditional one-dimensional array. The results in this work (see Sec. III B) suggest that even covering the same flow area, arranging turbines in the vertical direction in addition to lateral can increase both the power coefficient and efficiency. 
We note that this new kind of array has been designed in Uldolmok tidal current power plant project (Fig. 2), nevertheless the efficient way to investigate its performance is still lacking. So, modelling two-dimensional arrays becomes the first motivation of this work. 
\begin{figure}
  \centerline{\includegraphics[width=10cm]{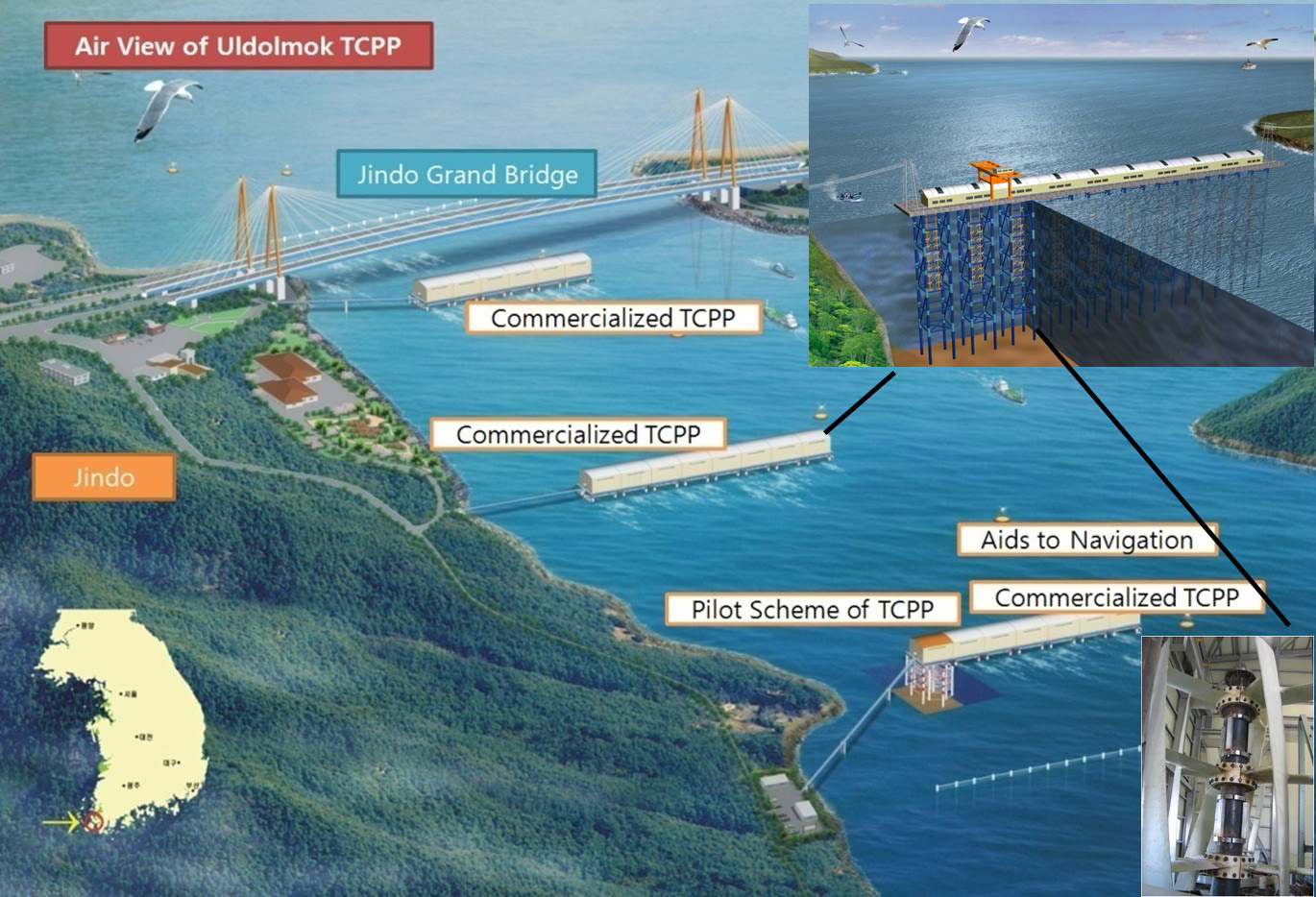}}
  \caption{Air view of Uldolmok tidal current power plant adapted from \citet{Li2014}.}
\end{figure}

At the same time, partial array models derived by \citet{Nishino2012, Nishino2013} do not account for the deformation of free surface. 
Depending on Froude number, the depth reduction can exact an influence on the power coefficient, efficiency and optimal operating condition of turbines. What's more, the decline of the free surface generated by energy extraction should also be considered in large scale channel model rather than turbine forces alone (for example, the surface change can be directly used to 1-D shallow water model with no approximation of the elevation change).
Usually the head loss amplitude is small, \citep{Vennell2015} so the change in the free surface can affect the whole dynamical system. Combining the free surface model with the NW12 model, \citet{Vogel2016} try to develop a one-dimensional array framework to address this problem. Considering the importance of the free surface effects, thus another motivation for this work is to include them on the partial array model.

Derived by these motivations, in this paper we extend the partial array model by allowing the vertical arrangement of turbines and depth change to investigate the performance of two-dimensional arrays in free surface flow. 
Compared with the rigid lid model, the new model can be used to describe more general conditions by adjusting Froude number. 
In vertical direction, the center section of the water offers the most tranquil conditions which are suitable for tidal turbine devices as the turbulence is strong near the free surface and sea bed. \citep{Myers2010} Considering this, we assume the array is still a partial fence in the vertical direction to avoid the upper and bottom flow. 
Usually the width of the channel is much larger than the depth, so the vertical confinement between the free surface and bottom is far greater than lateral boundaries. 
Based on this, here we make an important assumption that the vertical mixing of array scale flow takes place much faster than lateral mixing. 
Therefore, other than local and array scale, a new level of scale separation, vertical scale, is introduced. And this three-scale free surface model can be easily simplified to two-scale model (similar to the NW12 model however considers the free surface effects) to represent the scenario of one-dimensional arrays.

Directly extended from NW12, \citet{Cooke2016} raised a concept of sub-array by splitting a long row into several smaller sub-fences. Also, an additional sub-array scale is created other than local and array scale. For this work, the added scale is not from the interaction between different areas, but the asymmetry of wake mixing of one blockage area itself. The existing actuator disc theory is one dimensional, which is usually used to model the actuator disc in a flow with square section or rectangle section which width and height are in the same order. For a disc in a channel with far larger width than height, the mixing in vertical direction and lateral direction is no matter at the same time. This can be seen from the \citet{Nishino2012a} for a channel with an aspect ratio 4 and the limited power increase just benefits from this condition. 
For a high aspect ratio plate (representing two dimensional array after turbine scale mixing) in such a channel, if it is long enough in lateral direction, then every part can been seen as experience the same vertical mixing constrained by adjacent vertical strips (the adjacent strips are just like walls, the condition which has been modeled and verified using an experiment by \citet{Whelan2009}). So it is reasonable to assume the two-scale separation of wake for this kind of blockage area. Our numerical simulation (see Sec. IV) suggests that this assumption is reasonable. The key point is using this two-scale theory (essentially one-dimensional model) to describe the two-dimensional wake mixing of one blockage area.

It should be stressed that as with the NW12 model, the tidal array in this study still has one partial fence. For more complicated scenario, the present two-dimensional array can be arranged in rows along flow direction (thus becoming three-dimensional array). Besides the condition that two arrays are far away from each other to escape the array scale interference, the present model need to be improved greatly to represent three-dimensional arrays. However, of importance here is that one fence is preferred if we have finite number of turbines as \citet{Draper2014} showed that putting staggered turbines in the same fence is recommended from both the power coefficient and efficiency perspectives. 

As we discussed above, channel dynamics are ignored in previous partial array models. \citep{Nishino2012,Nishino2013,Cooke2016,Vogel2016} At the same time, existing studies describing the dynamics of the whole channel only concern macro-arrangement of the array across the whole width. \citep{Vennell2015} This raises questions that whether the benefits from certain arrangements still works and if yes, to what extent. To answer these questions, a necessary work is to combine these two aspects together. The effect of the channel dynamics on two-dimensional array would be analyzed in Sec. V. In this study, we still mainly focus on the dynamics within the array. More systematic analyses about the dynamics of the whole channel with different array configurations would be submitted soon. 

\begin{figure}
  \centerline{\includegraphics[width=13cm]{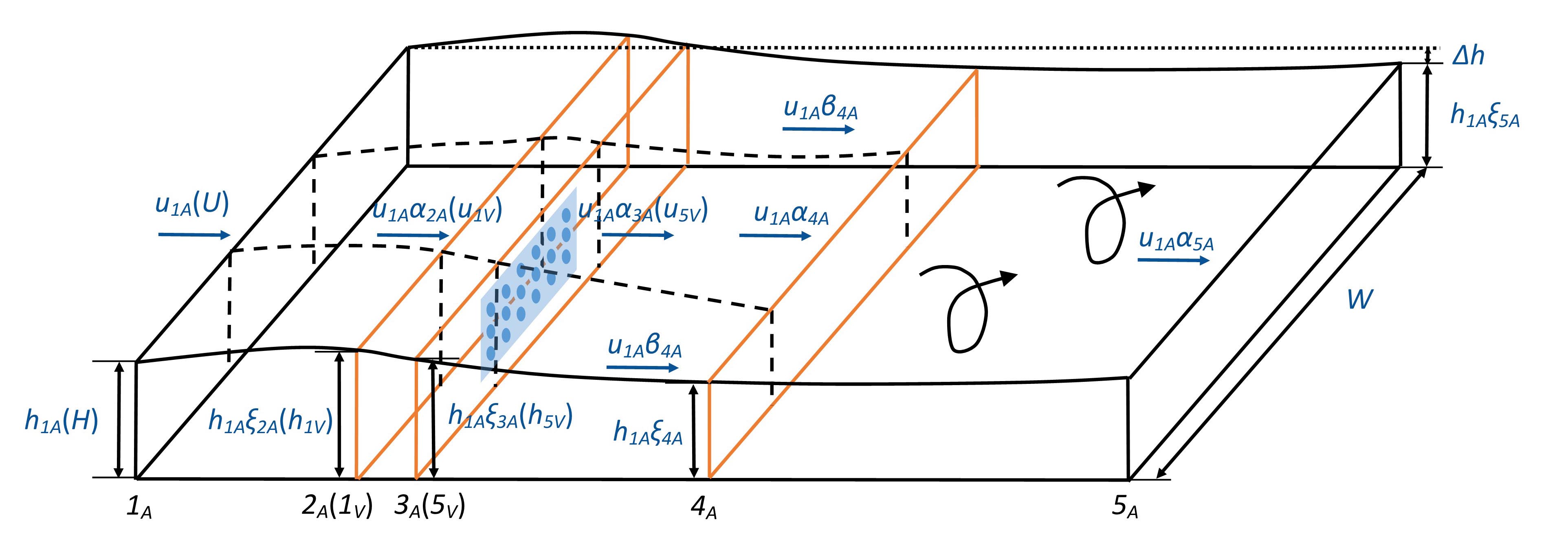}}
  \caption{Schematic of the whole model (mainly view of array scale flow). Turbines are arranged in a plane between $2_{A}$ and $3_{A}$.}
\label{fig:ka}
\end{figure}

\section{Mathematical Model}
Neglecting the channel dynamics, consider a large number of turbines which operate in exactly the same fashion partially span a rectangular channel of uniform velocity $U$, width $W$ and depth $H$ (Fig. 3). Along the flow direction, the deformation of the surface is allowed, and the channel is long enough so that the wake mixing is completed. Turbines are homogeneously arrayed in both the lateral and vertical directions. 

As with the NW12 model, there will be a scale separation between the flow around each device confined by other turbines surrounding it and that around the array restricted by channel boundaries.  
For two-dimensional arrays, additionally we assume a vertical scale separation besides local and array scales because the vertical contraction between the free surface and bottom is far beyond than lateral between channel banks. The vertical scale flow passage covers the whole height of flow (Fig. 4). In lateral direction, as will be discussed later, we makes no specific assumption about the width of vertical scale passage.  Note that the whole passages of local scale flow form the fence in vertical scale problem, and similarly for array scale problem, passages of vertical scale flow form the array scale fence.                                                                                                                                                                                                                                                                                                                                                                                                                                                                                                                                                                                                                                                                                                                                       
\begin{figure}
  \centerline{\includegraphics[width=10cm]{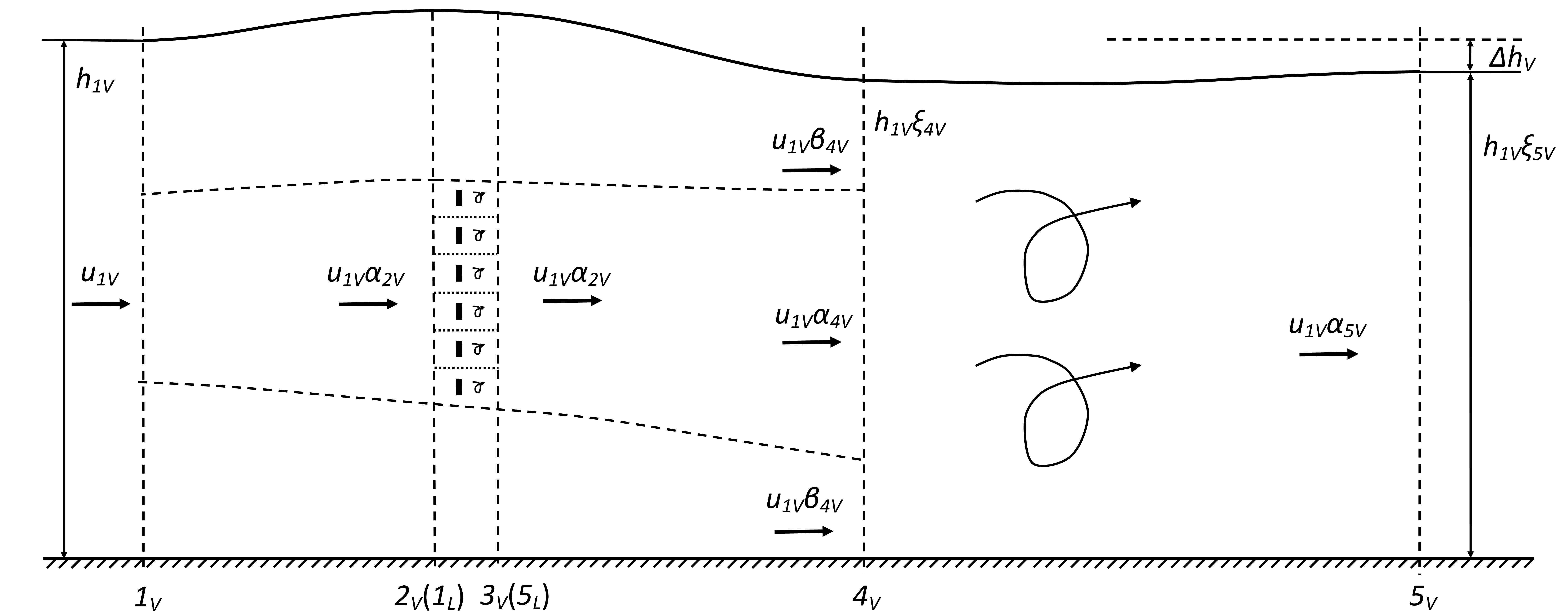}}
  \centerline{(a)}
  \centerline{\includegraphics[width=10cm]{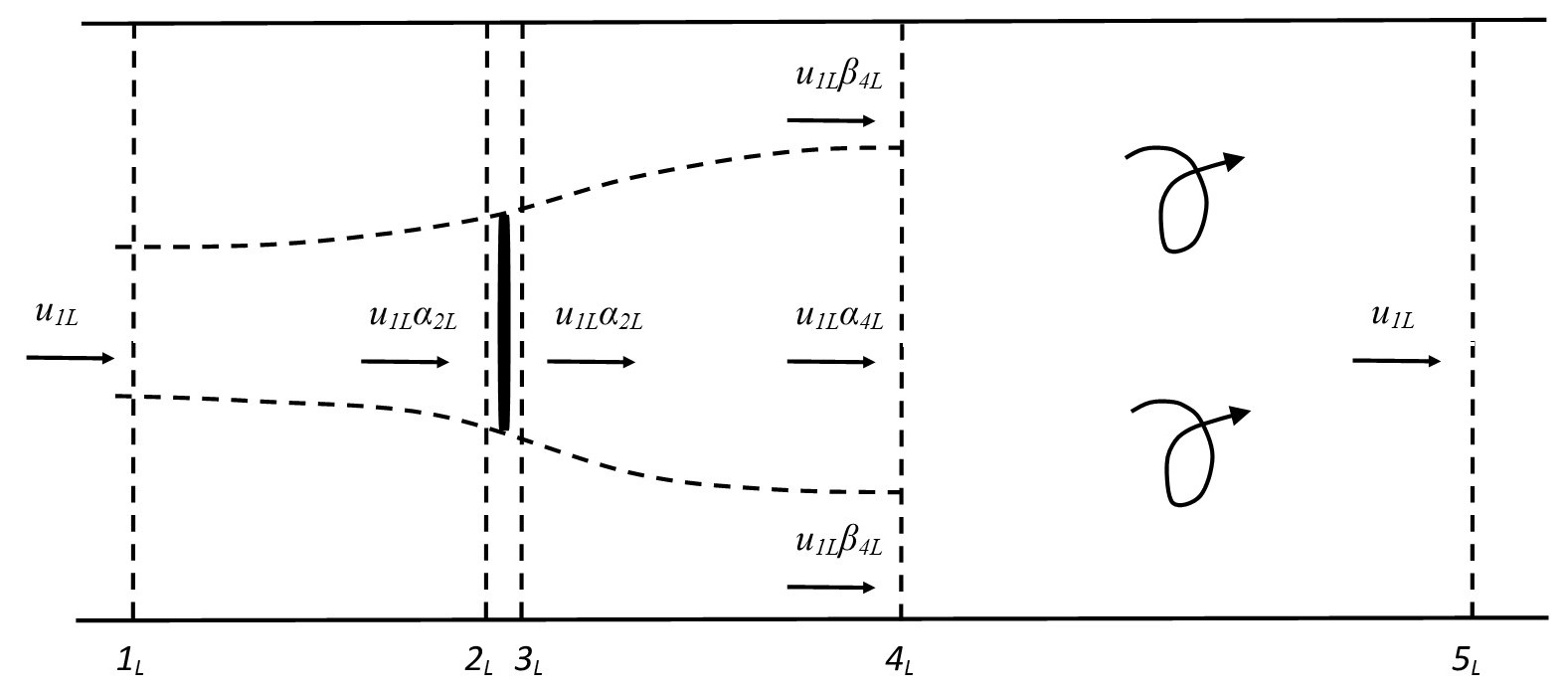}}
  \centerline{(b)}
  \caption{ Schematic of the vertical scale and local scale models: (a) vertical scale model. The fence composed of passages of local flow can be seen as a hypothetical turbine because the frontal and posterior cross-sectional area of local flow passage is the same. (b) local scale model. The flow passage is assumed to be constrained by constant boundaries.}
\label{fig:ka}
\end{figure}

Based on the linear momentum actuator disc theory, the turbine is seen as a momentum sink in this work. The whole channel loses energy only in discontinuous static head across turbines and mixing zones at local, vertical and array scales. Still five stations in each scale are defined. Station 1 and 5 represent the far upstream and downstream locations, where pressures and velocities are uniform across the cross-section of each scale problem. Stations 2 and 3 are immediately upstream and downstream of turbines or fences. Along the flow direction, pressure will equilibrate between bypass and core flows at station 4, however the velocity is different as the stream tube expands between station 1 and 4.  Wake mixing of all three flows are still considered to occur only after downstream of pressure equilibrating location station 4 and be totally completed before far downstream location station 5 (which means far-wake mixing alone). Pressures at station 1, 4 and 5 in vertical and array scale models are assumed as hydrostatic. 

In this three-scale model, we assume that the difference between characteristic lengths of three scales, turbine diameter, array height 
and array width is sufficient large, which requires large number of turbines or a long array in \citet{Nishino2013} (approximately ten turbines if we allow a five percent  difference compared with an array of forty or more turbines) in both the vertical and lateral directions. Thus the mixing sequence in time is that the vertical scale mixing takes much slower than local scale while much faster than array scale. And events in smaller scales also occur faster than the expansion of larger scale flows. In space, firstly this means that the core flow and bypass flow of smaller scale problems are completely mixed before the pressure equilibrium location of larger scale problems. Thus the direct interaction of the local, vertical and array scale wake mixing is neglected considering that only far-wake mixing is allowed. Secondly, more accurately the wake mixing of smaller scale problems is completed before the initiating of larger scale flow expansion, which means flow passages of local and vertical scale are constant in lateral direction. Again, the end effects are neglected in this work for a long array.

Particularly, for convenience we also make an assumption that besides expansion effects, the influence of gravity on the geometry of local scale flow passage is also ignored. Therefore, in vertical direction the local scale flow is also approximately considered to be confined by constant boundaries, which is the same with the flow in lateral direction. Under these conditions the local scale flow passage  has a rectangular uniform cross-section whose length side is the distance between cores of turbines, making it possible to be analysed by GC07 rigid lid model. Here it should be noted that this does not contradict with the fact that the whole problem is evaluated as the open channel flow. The flow volume change is allowed between station 2 and station 3 in vertical scale model and can be considered as occurring out of local scale flow passages.

\subsection{Basic formulations}

Before deducing the mathematical model, parameters should be explained clearly. Here we adopt the following dimensionless parameters in the process of modelling. The non-dimensional velocities for core and bypass flows are described using parameters $\alpha_{iL,V,A}$  and $\beta_{iL,V,A}$, respectively,  representing the ratio of the core and bypass velocities to upstream velocities in each scale $u_{1L,V,A}$. Subscripts $L$, $V$ and $A$ are, respectively, introduced to denote the local scale, vertical scale and channel scale problems, while subscript numbers from 1 to 5 represent the station of these parameters. 
For free surface flows, the non-dimensional height $\xi_{iV,A}$, which are the ratio of the surface heights in different stations $h_{iV,A}$ to reference height, i.e. the upstream heights in each scale $h_{1V,A}$, are also involved. 

Extended from the NW12 model, the larger scale problem provides upstream conditions for smaller scale problem, such as the velocity and the pressure. In turn,  the downstream conditions of smaller scale problem provide conditions for larger scale wake problem, which means that station 2 and 3 of the larger scale are, respectively, the station 1 and station 5 of the smaller scale. This fact is named as continuity conditions for these three different scale problems. 

Due to this, we have
\begin{align}
u_{2A}=u_{1V},u_{3A}=u_{5V}, \\
u_{2V}=u_{1L},u_{3V}=u_{5L},
\end{align}
where all velocity's symbols $u_{iL,V,A}$ are core flow velocities for station 2 and 3 unless specified emphasis. 

Besides, for open channel problem, the elevation head (pressure)  and the surface height should also satisfy continuity conditions. For elevation head that means
\begin{align}
z_{2A}=z_{1V},z_{3A}=z_{5V}, \\
z_{2V}=z_{1L},z_{3V}=z_{5L},
\end{align}
while for the surface height we have 
\begin{equation}
h_{2A}=h_{1V},h_{3A}=h_{5V}.
\end{equation}

The continuity conditions Eqs. (1)-(5) which couple the three problems are naturally encapsulated in the theoretical model. Physically, considering the turbine thrust is the only external force in the whole flow domain, the dynamical condition requires that the total thrust at all scales must be the same. The key of the present model is that these two coupling conditions, i.e. the continuity condition and dynamical condition, link the local, vertical and array scale problems together.  

As the basic coupling conditions have been introduced, in the sequel, we give the definitions of blockages, thrust coefficients and power coefficients.
Firstly, blockages in each scale which describe the confinement of flow are defined. The local blockage is the ratio of turbine area to cross-sectional area of the local scale passage:
\begin{equation}
B_{L}=\frac {\frac 14  \pi d^2}{(b+d)(s+d)}=\frac {\frac 14  \pi }{(\frac bd+1)(\frac sd+1)},
\end{equation}
where $d$ is the diameter of turbines, $s$ and $b$ are the vertical and lateral intra-turbine spacings, respectively. Meanwhile, vertical blockage and array blockage are defined as the frontal area of vertical and array scale fence normalized by upstream cross-sectional area of flow passage in each scale:
\begin{gather}
B_V=\dfrac {m(s+d) w_V}{h_{1V} w_V }=\dfrac {m(s+d)}{h_{1A}}    \dfrac{h_{1A}}{h_{2A}}=\dfrac {m(s+d)}{h_{1A}} \dfrac {1}{\xi_{2A}},\\
B_A=\dfrac {n(b+d) h_{2A}}{h_{1A}W}=\dfrac {n(b+d)}{W} \dfrac{h_{2A}}{h_{1A}}=\frac {n(b+d) }{W} \xi_{2A},
\end{gather}
where $w_V$ is the width of the vertical scale flow passage, $m$ and $n$ are, separatively, the number of turbines in vertical direction and lateral direction. Note that the finial expression of the vertical blockage is independent of $w_V$,  corresponding to the statement mentioned in Sec. II A that the width of the vertical scale flow passage is not specified. Here we can assume that laterally there is only one turbine in the vertical scale flow passage (which means $w_V=b+d$) to simplify the derivation. It's also worth noting that these two blockages depend on the unknown quality $\xi_{2A}$, so it is inconvenient to describe the arrangement of arrays by using them. In view of this we define two new blockages, namely designed vertical blockage and designed array blockage: 
\begin{gather}
B_{VD}=\frac {m(s+d)}{h_{1A}} =\frac {m(s+d)}{H}=\frac{1+\frac sd}{\frac {H}{md}},\\
B_{AD}=\frac {n(b+d)}{W}=\frac {1+\frac bd}{\frac {W}{nd}}.
\end{gather}
Rather than those original blockages ($B_V$ and $B_A$), designed blockages are depend only on the given parameters and can be used to measure the arrangement of arrays. Using Eqs. (9) and (10), vertical and array blockages are redefined as: 
\begin{equation}
B_V=B_{VD}  \frac {1}{\xi_{2A}} ,      B_A=B_{AD} \xi_{2A}.
\end{equation}

In addition, one can introduce the global blockage to describe the direct blockage of devices to the whole channel in the following way: 
\begin{equation}
B_G=\dfrac{mn \dfrac14 \pi d^2}{WH}=B_L B_V B_A.
\end{equation}

Next, we define three thrust coefficients in the local, vertical, and array scales: 
\begin{gather}
C_{TL}=\dfrac {\text{thrust on a single device}}{\text{upstream dynamics pressure} \times \text {single device area}}=\dfrac {T_L}{\dfrac 1 2 \rho u_{1L}^2  \dfrac 14 \pi d^2 }, \\
C_{TV}=\dfrac {\text{thrust on the vertical scale problem}}{\text{upstream dynamics pressure} \times \text{fence area}}= \dfrac {T_V}{\dfrac 12 \rho u_{1V}^2 m (s+d)(b+d) },\\
C_{TA}=\dfrac {\text{thrust on the array scale problem}}{\text{upstream dynamics pressure} \times \text{fence frontal area}}=\dfrac {T_A}{\dfrac 12 \rho u_{1A}^2  n(b+d) h_{2A} }.
\end{gather}

By applying continuity conditions Eqs. (1)-(5) and dynamical conditions ($T_V=mT_L$ and $T_A=nT_V$), the above three thrust coefficients have the relationship:  
\begin{gather}
C_{TV}=C_{TL} B_L \alpha_{2V}^2,\\
  C_{TA}=C_{TV} B_V \alpha_{2A}^2.
\end{gather}
These two equations will be used as coupling equations to link three problems together in Sec. II B.

For power coefficients, as we only concern the power which can be converted into electricity, the numerator of the power coefficient should be the power extracted by turbines (There is another dimensionless parameter $\eta$ to measure the dissipated energy.). Based on the two coupling conditions, three different power coefficients, namely local, global and channel coefficients are defined as: 
\begin{gather}
C_{PL}=\dfrac {P_L}{\dfrac 12 \rho u_{1L}^3  \dfrac 14 \pi d^2 }=C_{TL} \alpha_{2L},
\\
C_{PG}=\dfrac {nmP_L}{\dfrac 12 \rho u_{1A}^3 nm \dfrac 14 \pi d^2 }=C_{PL} \alpha_{2V}^3 \alpha_{2A}^3,
\\
C_{PC}=\dfrac {nmP_L}{\dfrac 12 \rho u_{1A}^3 WH}=C_{PL} B_L B_V B_A \alpha_{2V}^3 \alpha_{2A}^3.
\end{gather}

Results from both the analytical model and numerical simulations show that the force of partial array decreases the upstream velocity of the local scale problem, so the power coefficient of local scale is no longer an appropriate measure of turbines. Instead, global power coefficient $C_{PG}$ is more accurate to represent the performance of the turbine in an array. High global power coefficient means a turbine with certain area can gain more energy from channel. Besides $C_{PG}$, sometimes we also concern the total power we can gain from the channel. The channel power coefficient $C_{PC}$ defined in Eq. (20) is a way to measure it which relates to the number and the area of turbines in addition to $C_{PG}$.

\subsection{Governing equations of the whole problem} 
In this subsection, we will derive the governing equations of the three-scale model. Firstly, we consider the local scale problem. As shown in the work of \citet{Garrett2007} to analyse the flow field confined by rigid boundaries, the conservation of mass, momentum and energy in the near field flow (station 1 to 4) give us: 
\begin{equation}
C_{TL}=(1-\alpha_{4L}) \frac{((1+\alpha_{4L})-2B_L \alpha_{2L})}{(1-B_L  \alpha_{2L} / \alpha_{4L} )^2 }.
\end{equation}
in which the two non-dimensional velocities $\alpha_{2L}$ and $\alpha_{4L}$ satisfy:
\begin{equation}
\alpha_{2L}=\dfrac {1+\alpha_{4L}}{(1+B_L)+\sqrt{(1-B_L )^2+B_L (1-1/\alpha_{4L} )^2 )}}.
\end{equation}

For vertical scale problem, as velocities and areas of the front side and back side of the fence are the same, the fence which created by local flow passages now becomes a hypothetical turbine (which can be modeled as a momentum sink in actuator disk theory). So the vertical scale problem can be directly modelled by \citet{Houlsby2008} or \citet{Whelan2009}.
 
More precisely, starting from the mass conservation along the vertical scale flow passage, $\alpha_{2V}$ 
can be obtained by: 
\begin{equation}
\alpha_{2V}=\frac {\alpha_{4V} (\beta_{4V} \xi_{4V}-1)}{B_V (\beta_{4V}-\alpha_{4V} ) }.
\end{equation}
Applying Bernoulli equation along the bypass flow, equation (23) gives: 
\begin{equation}
\alpha_{2V}=\frac {\alpha_{4V} (\beta_{4V} (1-\dfrac {Fr_V^2}{2} (\beta_{4V}^2-1))-1)}{B_V (\beta_{4V}-\alpha_{4V} ) },
\end{equation}
where $Fr_V=u_{1V}/\sqrt{gh_{1V} }$. Similar to the case of vertical blockage, $Fr_V$ is still unknown but can be expressed by channel Froude number $Fr=U/\sqrt{gH}$, the known parameter for a certain channel, and the unknown parameters $\alpha_{2A}$ and $\xi_{2A}$ in the form: 
\begin{equation}
Fr_V=Fr\alpha_{2A} \sqrt\frac {1}{\xi_{2A}}.
\end{equation}

Next, using Bernoulli equation to measure the pressure drop across the hypothetical turbine yields the expression for vertical scale thrust: 
\begin{equation}
T_V=\dfrac 12 \rho u_{1V}^2 m(s+d) w_V (\beta_{4V}^2-\alpha_{4V}^2 ).
\end{equation}

Then, the thrust coefficient is a function of the non-dimensional wake velocity:
\begin{equation}
C_{TV}=(\beta_{4V}^2-\alpha_{4V}^2 ).
\end{equation}

Note that the thrust can also be determined by applying the conservation of momentum for near field flow:
\begin{equation}
T_V=w_V (\dfrac 12 \rho g(h_{1V}^2-h_{4V}^2 )+\rho u_{1V}^2 h_{1V} (1-B_V \alpha_{2V} )(1-\beta_{4V} )+\rho u_{1V}^2 h_{1V} B_V \alpha_{2V} (1-\alpha_{4V} )).
\end{equation}

Combining it with Eq. (26) to eliminate $T_V$,  and making use of  Eq. (24), we can obtain a quartic expression of $\beta_{4V}$ in terms of $B_V$, $\alpha_{4V}$ and $Fr_V$
\begin{gather}
Fr_V^2 \beta_{4V}^4+4 \alpha_{4V} Fr_V^2 \beta_{4V}^3+(4B_V-4-2Fr_V^2 ) \beta_{4V}^2+(8-8\alpha_{4V}-4Fr_V^2 \alpha_{4V} ) \beta_4V \notag\\
+(8\alpha_{4V}-4+Fr_V^2-4\alpha_{4V}^2 B_V )=0.
\end{gather}

So far, we have derived equations of the local scale and vertical scale problems. Considering the full recovery of the wake in local scale flow, these two problems can be modelled separately and then be linked by the coupling equation. 
To connect them, we make use of the relation between  $C_{TL}$ and $C_{TV}$  deduced by continuity and dynamical conditions as Eq. (16) shows. 
In detail, for continuity conditions, the vertical scale problem provides the upstream conditions for local scale, and in turn local scale problem provides conditions for vertical scale wake. As a result, applying conservation of momentum to the whole passage (station 1 to 5) of the local scale flow, we can write: 
\begin{equation}
T_V=mT_L.
\end{equation}
It should be noted that this equation actually reflects the dynamical condition. 
So for connection between the local and vertical scale problems, the dynamical condition can also be derived mathematically by using continuity condition, which is also true for NW12 model. 

Similarly, the downstream condition of the vertical scale problem used in solving array scale wake is obtained by applying the conservation of momentum for the whole vertical flow passage: 
\begin{equation}
\frac 12 \rho g w_V (h_{1V}^2-h_{5V}^2)-T_V=\rho w_V ( h_{5V} u_{5V}^2-h_{1V} u_{1V}^2).
\end{equation}
Following \citet{Houlsby2008}, one can rewrite Eq. (31) in the non-dimensional form that: 
\begin{equation}
(\xi_{1V}-\xi_{5V})^3-3(\xi_{1V}-\xi_{5V} )^2+(2-2Fr_V^2+C_{TV} B_V Fr_V^2)(\xi_{1V}-\xi_{5V})-C_{TV} B_V Fr_V^2=0.
\end{equation}

Now we derive the array scale problem equations. Considering the free surface effects, the extracting of energy results in difference of water depth and velocities between upstream and downstream of the vertical scale flow passage. This difference is also the velocities discontinuity across the array scale fence. 
Hence, the fence in the array scale problem is no longer a hypothetical turbine and the existing theory can not be used in this case. It forces us to rebuild the equations.

Inspired by the existing free surface model, \citep{Whelan2009, Houlsby2008} we try to use conservation of mass,  momentum and energy to get two expressions for thrust which described by two wake velocities, the core flow and bypass flow velocities, as well as the expression for velocity at the front side of fence.

Firstly, applying conservation of mass to the core flow leads to
\begin{equation}
u_{1A} h_{1A} A_{1A}=\alpha_{2A} \xi_{2A} u_{1A} h_{1A} L=\alpha_{3A} \xi_{3A} u_{1A} h_{1A} L=\alpha_{4A} \xi_{4A} u_{1A} h_{1A} A_{4A},
\end{equation}
where $A_{1A}$  and $A_{4A}$  are widths of stream tubes at station $1A$ and station $4A$, $L=n(b+d)$ is the width of fence. 

Meanwhile, it follows from the the conservation of mass for bypass flow that
\begin{equation}
u_{1A} h_{1A} (W-A_{1A})=\beta_{4A} \xi_{4A} u_{1A} h_{1A} (W-A_{4A}).
\end{equation}

Combining it with Eq. (33), we can get an expression for $\alpha_{2A}$:
\begin{equation}
\alpha_{2A}=\frac {\alpha_{4A} (\xi_{4A} \beta_{4A}-1)}{B_A (\beta_{4A}-\alpha_{4A})}.
\end{equation}

Next, we apply Bernoulli equation along the center streamline (representing the value of the core flow) and the free surface (representing the value of the bypass flow) to build the connection between water depth (pressure) and velocities. It yields 
\begin{gather}
\xi_{2A}=1-\frac 12 Fr^2 (\alpha_{2A}^2-1),\\
\xi_{3A}=\xi_{4A}-\frac 12 Fr^2 (\alpha_{3A}^2-\alpha_{4A}^2 ),\\
\xi_{4A}=1-\frac 12 Fr^2 (\beta_{4A}^2-1).
\end{gather}

Now we seek to get the equation for the thrust of all turbines on the array. First, it can be obtained by considering the conservation of momentum between station $1A$ and $4A$ that
\begin{gather}
\rho u_{1A}^2 Wh_{1A}+\frac 12 \rho gh_{1A} Wh_{1A}=\rho \alpha_{4A}^2 u_{1A}^2 A_{4A} \xi_{4A} h_{1A} +\rho \beta_{4A}^2 u_{1A}^2 (W-A_{4A}) \xi_{4A} h_{1A}\notag\\+ \frac 12 \rho g \xi_{4A} h_{1A} W \xi_{4A} h_{1A}+T_A.
\end{gather}
Combined with Eqs. (33) and (34), and then normalized by dynamical pressure multiplying fence frontal area, Eq. (39) becomes:
\begin{equation}
1-\xi_{4A}^2-Fr^2 B_A C_{TA}=2Fr^2 (\alpha_{4A} (1-\xi_{4A} \beta_{4A} )+(\beta_{4A}-1)).
\end{equation}
Note that the non-dimensional heights $\xi_{2A}$, $\xi_{3A}$, $\xi_{4A}$ and non-dimensional velocities $\alpha_{2A}$, $\alpha_{3A}$ can be expressed by $\alpha_{4A}$ and $\beta_{4A}$ by using Eqs. (33)-(38). So there are three unknown parameters in Eq. (40), $C_{TA}$, $\alpha_{4A}$ and $\beta_{4A}$.  If one of the these three parameters are specified, we still need another equation which includes them to close the array scale model. 

Therefore, following the existing method,\citep{Whelan2009, Houlsby2008} we try to integrate the pressure drop across the array scale fence to get the second expression for thrust. However, the thrust derived in this way equals to
\begin{equation}
T_F=\frac 12 \rho g L H^2 (\xi_{2A}^2-\xi_{3A}^2),
\end{equation}
which is no longer the total thrust of turbines $T_A$ in virtue of contradicting with the dynamical condition $T_A=nT_V$ in Eq. (31). Consequently, $T_A$ can not be obtained by this argument and we cannot get the similar quartic polynomial equation of wake velocity coefficients as in vertical scale model.

Although Eq. (41) is unnecessary, the information from two sides of the array scale fence (which are also the upstream and down stream of the vertical scale problem) is still useful. To close this problem, we investigate the relationship of $C_{TA}$ and $C_{TV}$, Eq. (17). Furthermore, due to continuity conditions Eqs. (1)-(5), the Eq. (32) relating to vertical scale parameters can be rewritten as
\begin{gather}
(\xi_{2A}-\xi_{3A} )^3-3(\xi_{2A}-\xi_{3A} )^2 \xi_{2A}+(2\xi_{2A}-2Fr^2 \alpha_{2A}^2+C_{TA}  Fr^2 )(\xi_{2A}-\xi_{3A} ) \xi_{2A}\notag\\-C_{TA}  Fr^2  \xi_{2A}^2=0.
\end{gather}

Till now, we have established two thrust equations, Eqs. (42) and (40) for $\alpha_{4A}$, $\beta_{4A}$  and $C_{TA}$. 
Eq. (42) itself can be seen as a part of array scale equations considering it provides another mathematical expression for array scale thrust coefficient. Of course, coupling conditions between the vertical scale problem and array scale problem are embedded in it. In this sense, the vertical scale problem and array scale problem are not completely separated as the NW12 model suggested. Furthermore, the dynamical condition here is only a physical condition rather than the mathematical deduction. 

Overall, the whole equation system of the three-scale free surface problem includes five parts: 

(i) 	near field equations of local scale problem: Eqs. (21) (22);

(ii)	near field equations of vertical scale problem: Eqs. (24) (25) (27) (29);

(iii)  near field equations of array scale problem, Eqs. (33) (35) (36) (37) (38) (40) (42);

(iv)	coupling equation between the local scale problem and vertical scale problem: the relation between local scale thrust coefficient and vertical scale thrust coefficient Eq. (16) (naturally including the whole field equation of local scale) ;   

(v)	coupling equations between the vertical scale problem and array scale problem: the relation between vertical scale thrust coefficient and array scale thrust coefficient Eq. (17)

It should be noted that the above three-scale equation system which investigates the performance of the two-dimensional array can easily be simplified to model the one-dimensional array. For one-dimensional array, there are only two scale flows in model, i.e. the local scale flow and array scale flow, and the local scale flow is directly confined by channel bed and free surface. Thus the equation system of the one-dimensional array only includes part (ii), part (iii) and part (v). For comparison, the performance of the one-dimensional array will also be shown in Sec. III B, however we mainly focus on two-dimensional array in this work.

\section{Results}

\begin{figure}
{\includegraphics[width=7cm]{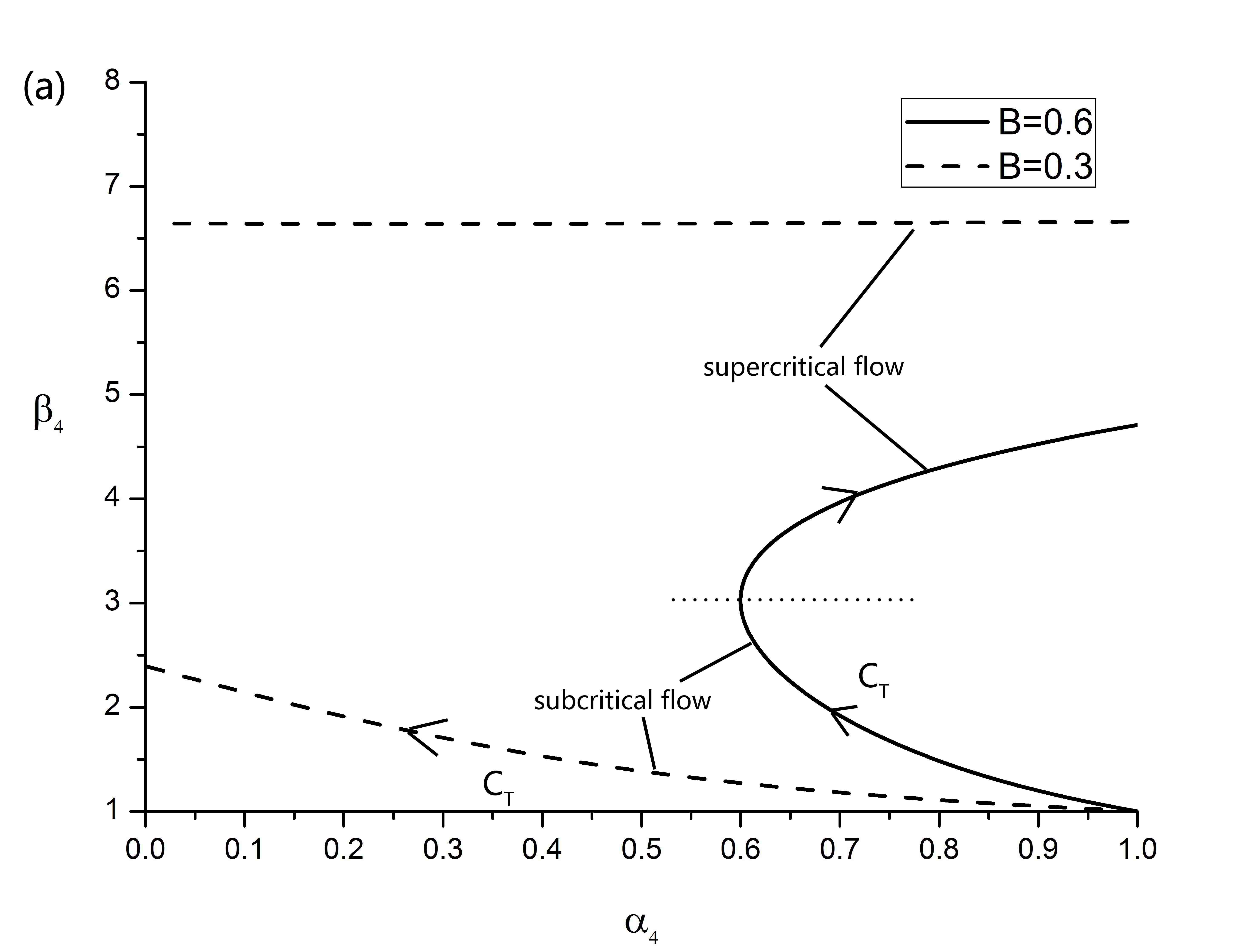}}
{\includegraphics[width=7cm]{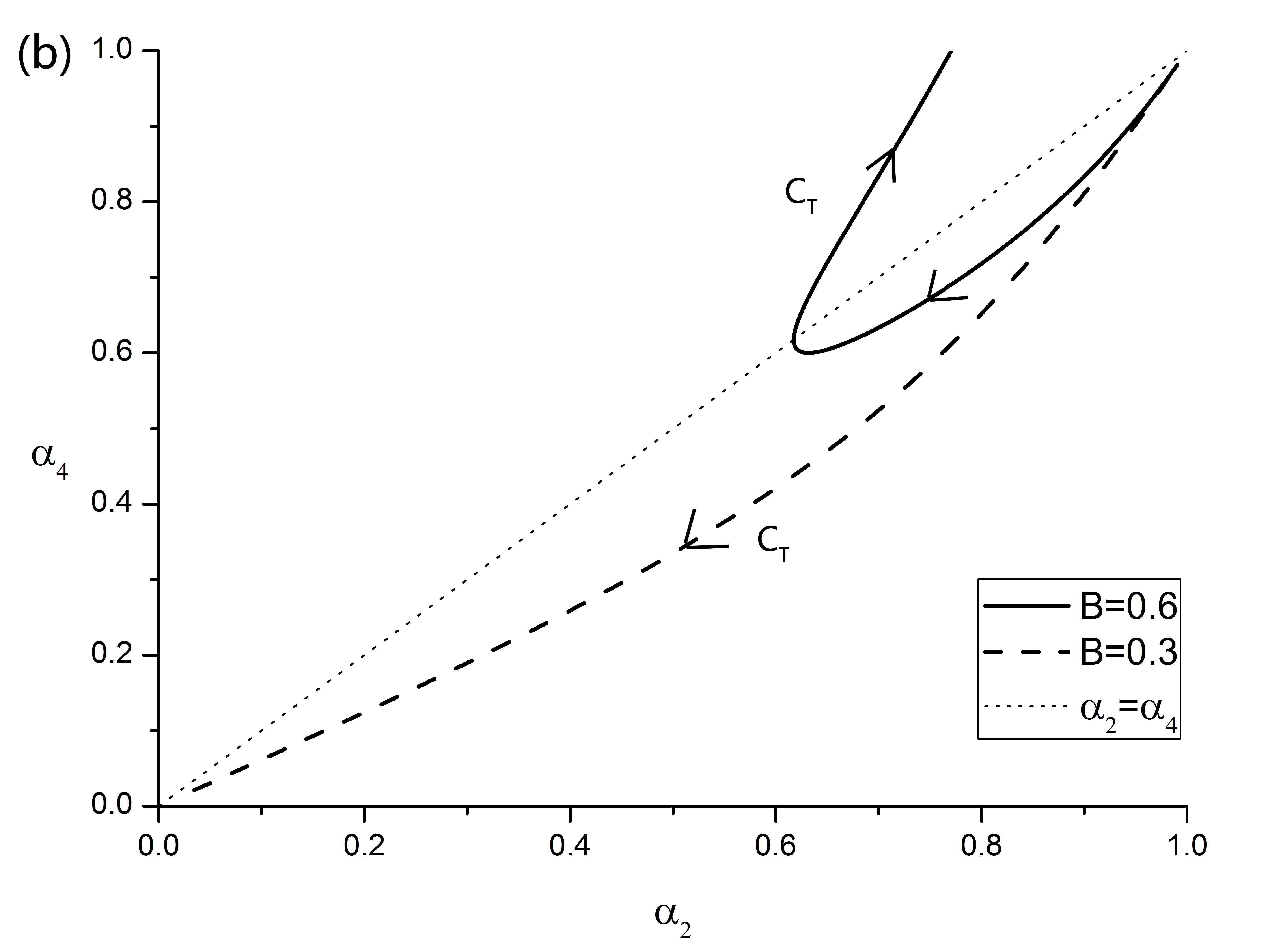}}
  \caption{The structure of solutions for the open channel model with $Fr=0.2$ used in the vertical scale problem. When the blockage effects are severe, a physically meaningful solution which satisfies the subcritical requirement (a) and slowing down requirement (b) only corresponds to a part of $\alpha_{4}$ and $\alpha_{2}$. }
\label{fig:ka}
\end{figure}
\begin{figure}
 {\includegraphics[width=7cm]{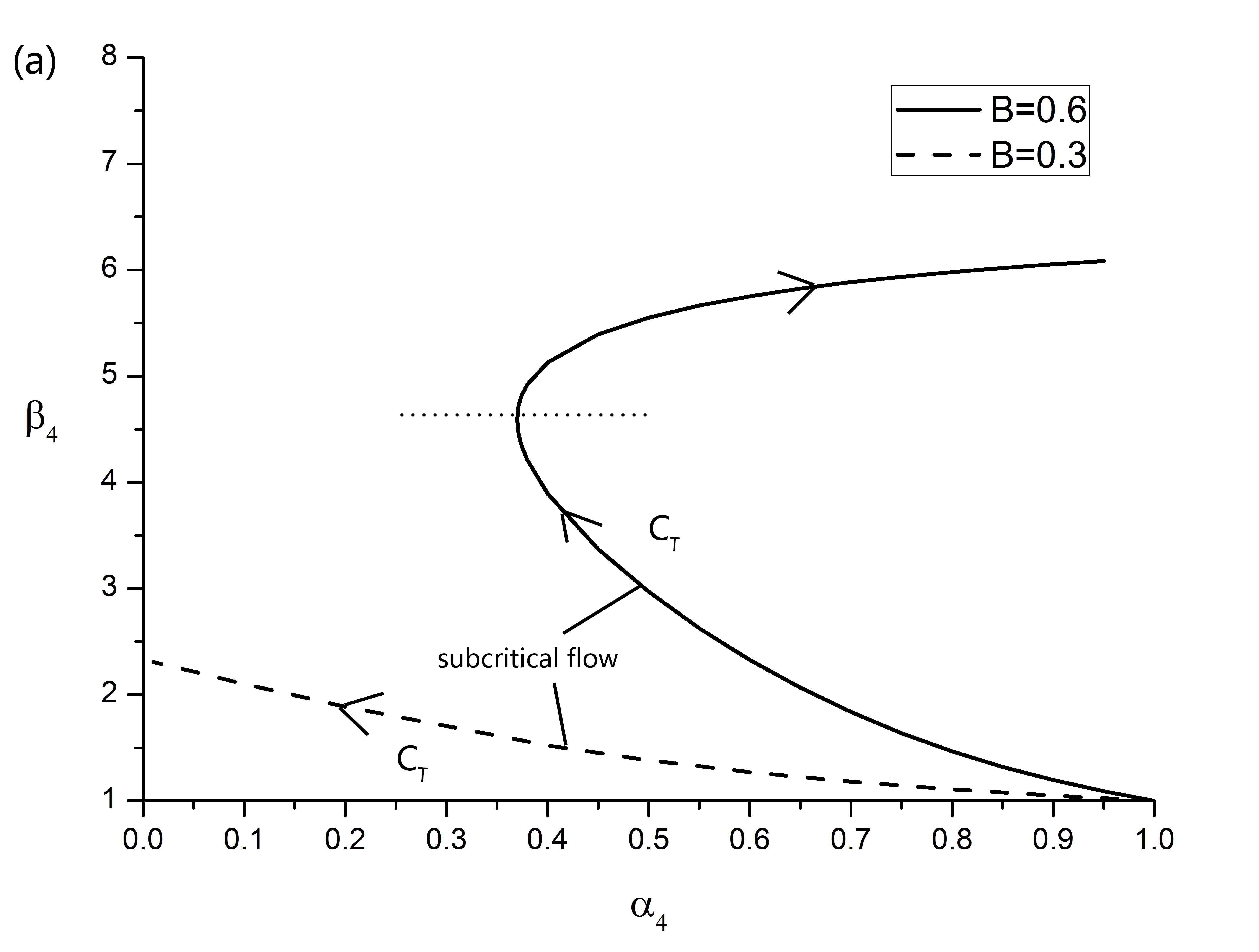}}
 {\includegraphics[width=7cm]{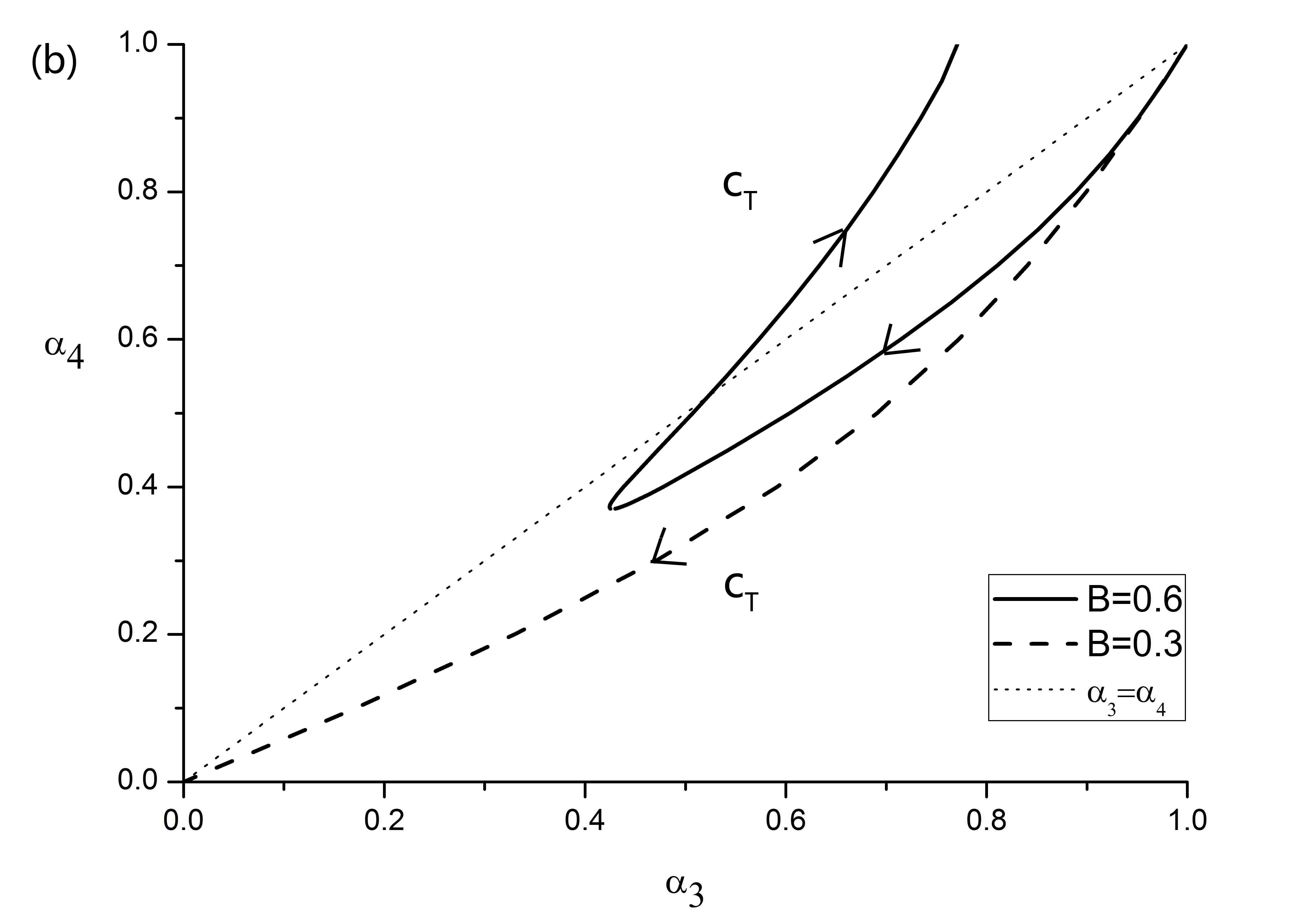}}
  \caption{The structure of solutions for open channel model with $Fr=0.2$ used in the array scale problem. The structure is similar with the existing open channel model in vertical scale problem except the replacement of $\alpha_{3}$ to $\alpha_{2}$ in (b) considering the decrease of water depth.}
\label{fig:ka}
\end{figure}

For this three-scale model, when one of the independent variables is determined, the whole equation system can be solved together by numerical method as long as it has a physically meaningful solution. Usually, for actuator disc theory the turbine induction factor $a_L=1-\alpha_{2L}$, a function of turbine operation, is specified.\citep[e.g.][]{Garrett2007, Nishino2012} However, decreasing the value of $\alpha_{2L}$ from 1 to 0, a physically meaningful solution does not always exist, especially for a system with high blockages and high Froude number. Compared with the rigid lid model, the solution of the free surface model is far more complex. 
For the open channel model used in vertical scale problem, \citep{Whelan2009,Houlsby2008} when the blockage effects are severe, only a part of $\alpha_{4}$ and $\alpha_{2}$ satisfy the subcritical requirement and slowing down requirement $0<\alpha_{4V}<\alpha_{2V}<1$ (Fig. 5). Compared with this, the array scale model has far more solutions, however the structure of the possible solutions is similar with the existing model (Fig. 6).

To get all possible values for this huge equation system is extremely difficult and time consuming. So here thrust coefficient $C_{TV}$ or $C_{TA}$ (directly linking the theoretical model with experiments using porous disks or actuator disk numerical simulations) is appointed. 
In this way the three-scale free surface problem can be solved separately and one by one starting from the array scale, then vertical and finally local scale problems. All  solutions in each scale can be obtained and analysed. 
Of importance here is that although the breakdown of solution is also existing in the array scale problem, the calculated results show that for the whole equation system, the vertical scale problem is responsible for the lacking of the physically meaningful solutions. 
Therefore, we focus on the vertical scale problem to choose physically meaningful solutions.

After getting solutions, the power coefficients can be obtained by using Eqs. (18-20). 
Besides power coefficients, the value of the efficiency and water depth change can also be calculated. The efficiency is especially important if the total potential energy which can be exploited is limited. Under such circumstances, high efficiency may be a preferred measure of turbines. Additionally, together with water depth change, the efficiency also represents the influence of deploying turbines on environment. Before deciding to build a tidal farm in certain channel, an assessment of environmental influence is necessary as arranging a large number of turbines may change the flow and water depth massively. 

Firstly, the momentum conservation of the whole channel between upstream and downstream of the array scale problem yields a cubic equation for the depth change:  

\begin{equation}
(\dfrac {\Delta h}{H})^3-3(\dfrac {\Delta h}{H})^2+(2-2Fr^2+C_{TA} B_A Fr^2)\dfrac {\Delta h}{H}-C_{TA} B_A Fr^2=0.
\end{equation}

Then the efficiency can be expressed as: 
\begin{equation}
\eta=\dfrac {\text{power extracted by turbines}}{\text{power removed in the channel}} =\dfrac {B_G C_{PL} \dfrac 12 \alpha_{2V}^3 \alpha_{2A}^3 Fr^2}{\dfrac{\Delta h}{H} (1-\dfrac {Fr^2  (1-\Delta h/2H)}{(1-\Delta h / H )^2})}.
\end{equation}

\subsection{Typical solutions}
\begin{figure}
  \centerline{\includegraphics[width=9cm]{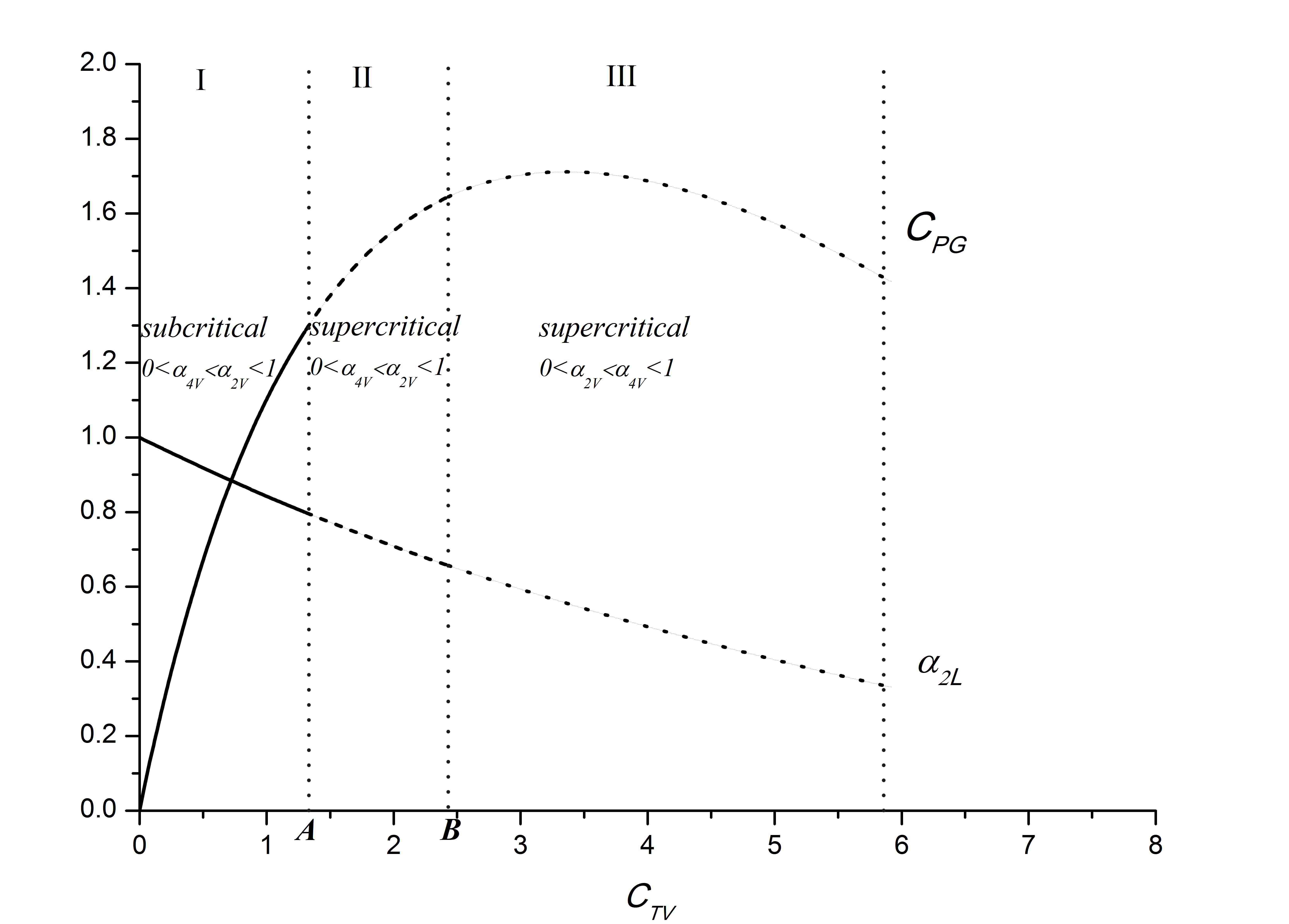}}
  \caption{The breakdown of the desired solutions for high blockages condition, $B_L=0.6$, $B_{VD}=0.8$ and $B_{AD}=0.7$, $Fr=0.2$. Solutions of zone I are physically meaningful for this model.}
\label{fig:ka}
\end{figure}

An example of solutions of the new model with high blockages ($B_L=0.6$, $B_{VD}=0.8$, $B_{AD}=0.7$ and $B_G=0.336$) is shown in Fig. 7 to illustrate the losing of physically meaningful solutions. Here the breakdown of lines (where $C_{TV}$  approaches to 5.9 or $\alpha_{2L}$ approximates 0.35) means all of the numerical solutions violate the basic requirement for wake velocity that $\beta_{4V}>1$ and $0<\alpha_{4V}<1$. Increasing the $C_{TV}$  from zero (which equals to decrease the $\alpha_{2L}$ from 1 to 0), results shown as solid line (zone I) mean that they meet the ‘subcritical’ requirement 
and slowing down requirement. Further increasing $C_{TV}$ (or decreasing $\alpha_{2L}$), stating from point $A$, results become to represent ‘supcritical’ flow. The dashed line (zone II) results still meet the slowing down requirement. However, after point $B$ (dotted line in zone III), the supercritical results become $0<\alpha_{2V}<\alpha_{4V}<1$. 
Here we choose the solution of zone I which satisfy the subcritical and slowing down requirements as physically meaningful one. Different with the complicated solution in this case, usually for blockages not close to 1, the desired solution (zone I) can cover most of the $\alpha_{2L}$ range including the maximum value of $C_{PG}$.
\begin{figure}
  \centerline{\includegraphics[width=9 cm]{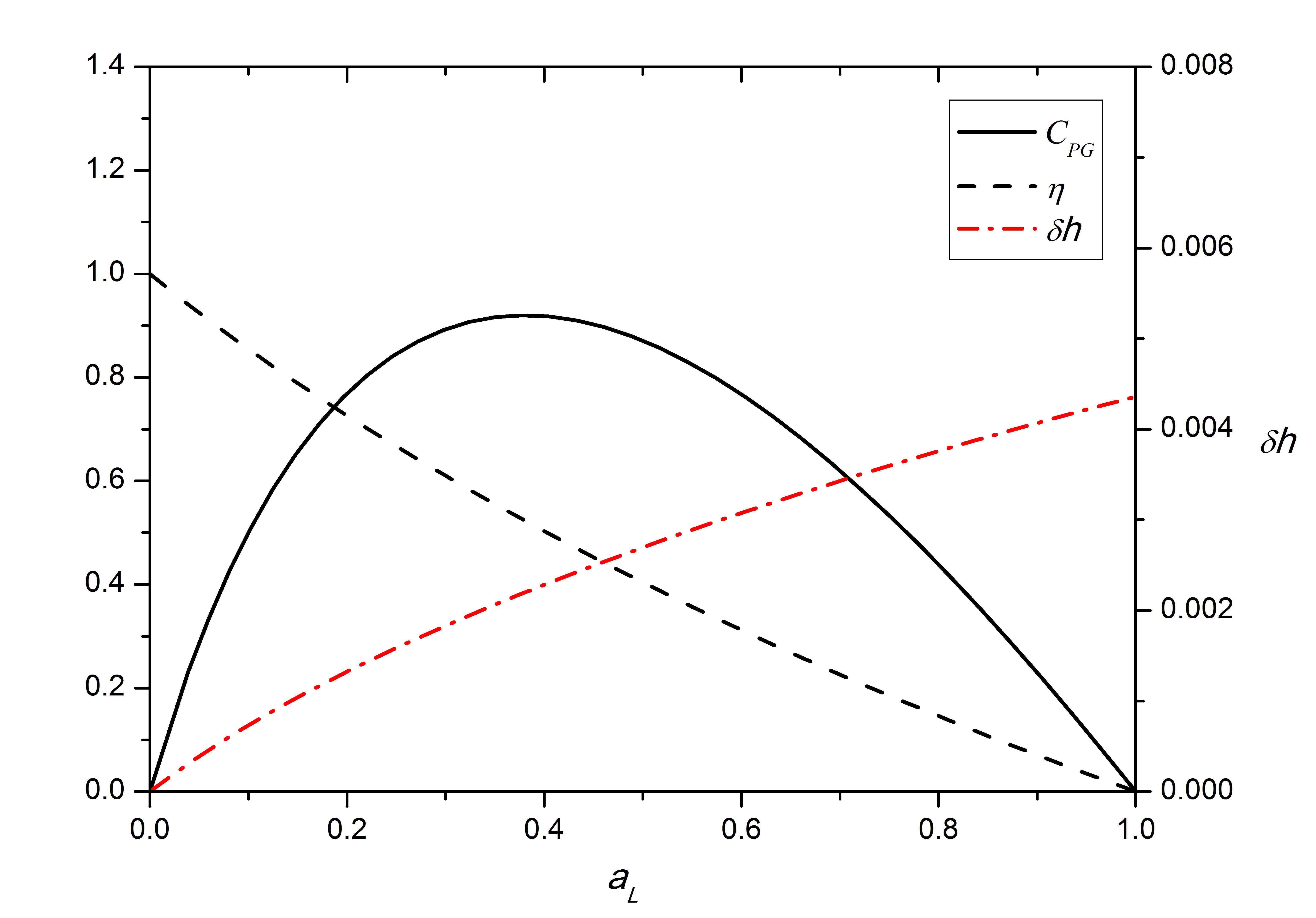}}
  \caption{Variation of  $C_{PG}$, $\eta$, and $\Delta h$ with $a_{L}$ ($B_L=0.4$, $B_{VD}=0.5$,$B_{AD}=0.3$ and $Fr=0.2$).}
\label{fig:ka}
\end{figure}

Fig. 8 explores the power coefficient, efficiency and depth change of normal blockages ($B_L=0.4$,$B_{VD}=0.5$ and $B_{AD}=0.3$). It should be noted that the original results 
along with  $C_{TV}$  have been transformed into varying with the local induction factor (which monotonically has the same trend with thrust coefficient). The maximum $C_{PG}$ occurs near $0.4$ for $a_{L}$ ($0.6$ for $\alpha_{2L}$) and exceeds the Lanchester Betz limit of 16/27. 
Compared with $C_{PG}$, the trend of $\eta$ and $\Delta h$ is simple. $\eta$ decrease monotonically with the increase of  $a_{L}$ while $\Delta h$ is the opposite. For a given $C_{PG}$ which is below the $C_{PGmax}$ in Fig. 8, two corresponding $a_{L}$ are founded. The smaller one is preferred as it means little environmental influence, lower structural load and high energy efficiency. A notable point is that the maximum $C_{PG}$ does not coincide with maximum efficiency and minimum depth change and thrust, so we need to balance the importance of these parameters. 
If we decrease some power from optimal operation (near $a_L=0.4$ in this case from a power coefficient perspective), the efficiency will increase in a larger amplitude. For example, the maximum power coefficient is 0.92 with efficiency equalling to 0.53, while the efficiency becomes 0.67 when the power coefficient decreases to 0.84 in $a_L=0.25$. 
The decrease of $a_L$ causes less velocity difference between the core flow and bypass flow, which is also true for the vertical and array scale problems. Thus less energy dissipated in the wake mixing process. 
Also, in this way the force exerted on turbines and the depth change is smaller. So a comprehensive assessment is needed to choose appropriate operations for turbines. In this work considering the features of the momentum model which represents the upper limit of power extraction, the power coefficient is emphasized. 

\subsection{Comparison between two-dimensional and one-dimensional arrays}

Now we consider the comparison of results (mainly maximum global power coefficient) 
between two-dimensional arrays (illustrated by the three-scale model) and one-dimensional arrays (illustrated by the two-scale model) to show the advantages of the vertical arrangement of turbines. 
For a relatively deep channel, assume that thirty percent of channel's width will be fully exploited (which means that the designed array blockage is fixed as 0.3). To utilize this part of channel, we can put one large turbine or a lot of small turbines in vertical direction. For comparison, the total area of turbines (which can be a measure of input or cost) in these two kinds of array are the same. Firstly we consider the one-dimensional array. If we only arrange one turbine in vertical direction, the local blockage is usually small. Practically for a 100 m depth channel, the diameter of turbine could be 50 m considering constraint from geography, flow features and manufacture. The local blockage is approximately 0.25 when the intra-turbine spacing is half the diameter. For two-dimensional array, considering the fixed global blockage, the local blockage multiplying the designed vertical blockage is 0.25, equalling to the local blockage of one-dimensional array. As $C_{PG}$ has the same trend with $C_{PC}$ for fixed global blockage, here we only show the results of the global power coefficient. 

\begin{figure}
  \centerline{\includegraphics[width=11cm]{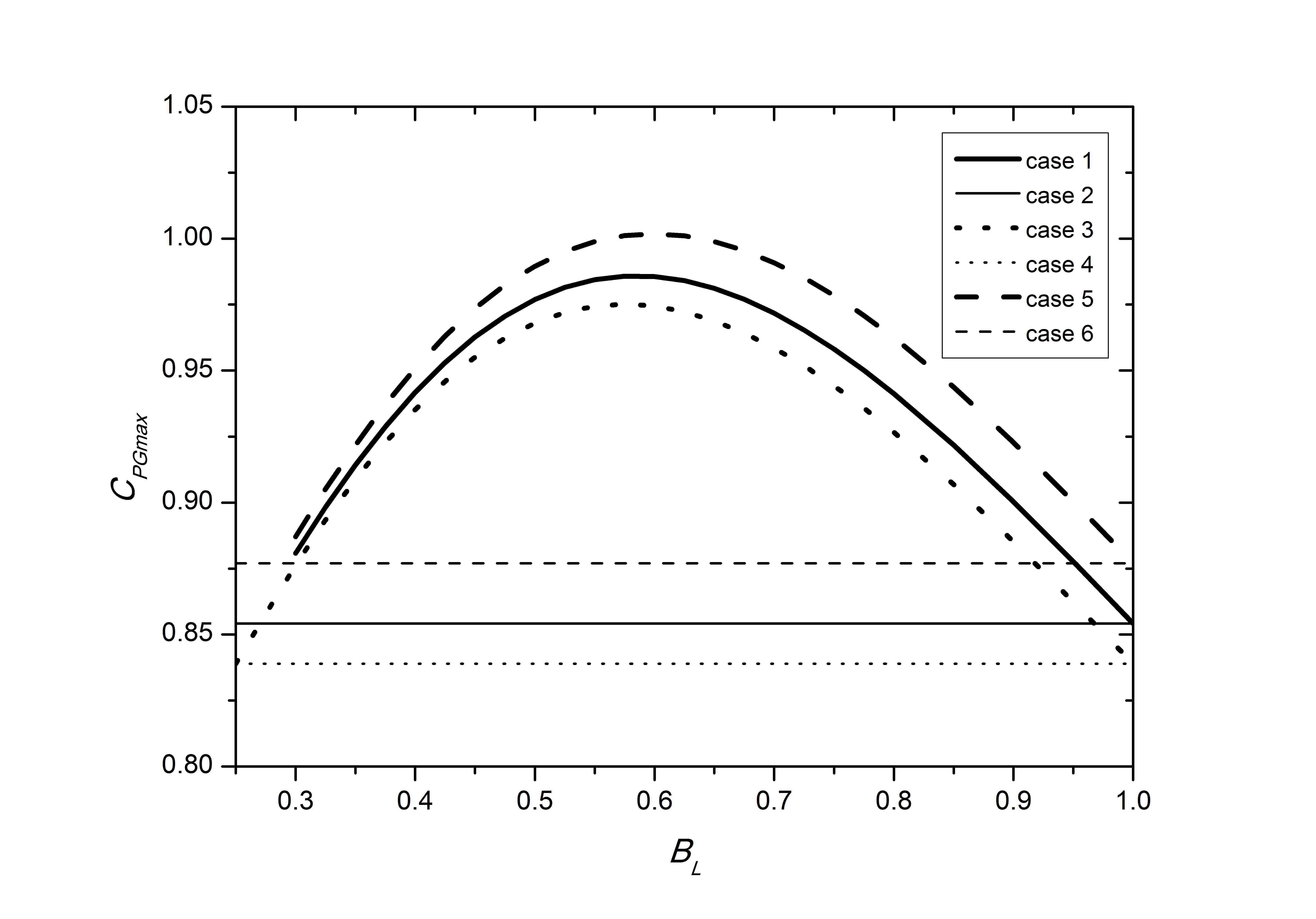}}
  \caption{Variation of $C_{PGmax}$ with local blockage for fixed $B_G=0.075$ and $B_{AD}=0.3$. Case 1: three-scale free surface model with $Fr=0.2$; case 2: two-scale free surface model with $Fr=0.2$; case 3: three-scale rigid lid model; case 4: two-scale rigid lid model (NW12); case 5: three-scale model with $Fr=0.3$; case 6: two-scale model with $Fr=0.3$.}
\label{fig:ka}
\end{figure}
The thick solid line (case 1) in Fig. 9 represents the maximum global power coefficient of the three-scale free surface model as a function of local blockage when Froude number is set to 0.2, whereas the thin solid line (case 2) describes the two-scale condition.  
As can be seen from the figure, the local blockage for the three-scale model can be adjusted to optimal value to get higher $C_{PGmax}$ from 0.854 in two-scale model to 0.986 in three-scale model where local blockage is approximately 0.6. 
Of importance here is that theoretically the two-scale free surface model is  incorporated within the three-scale model when $B_L=1$, $B_{VD}=0.25$. Under this condition the local scale flow separation disappear and the three-scale model naturally becomes two-scale. Easy to see that the result of $B_L=1$, $B_{VD}=0.25$ in the three-scale model is identical with that in the two-scale model. 
For $B_L=0.25$, $B_{VD}=1$, also the three-scale model becomes two-scale (The data of these condition do not shown in the figure for mathematical restrictions as it is hard to get the desired subcritical solution when vertical blockage approaches 1.). However, the local scale problem does not include the free surface effects in this scenario so that the results is slightly different with the value of case 2.

For comparison, also presented here are the results of the rigid lid model  (case 3 for the three-scale model and case 4 for the two-scale model) and free surface model when Froude number is 0.3 (case 5 for the three-scale model and case 6 for the two-scale model). It is worth noting that the results of the two-scale rigid lid model (case 4) actually correspond to the solution of the NW12 model. Meanwhile, the three-scale model is built by directly adding a scale to the NW12 model or substituting the near field equation (ii) and (iii) in this work by the GC07 model. It can be seen that the rigid lid model hold the smallest results and the free surface model with $Fr=0.3$ has much larger $C_{PGmax}$ than $Fr=0.2$, showing that the free surface effects increase the maximum power coefficient.
We also note that more increase is found from the model with $Fr=0.2$ to $Fr=0.3$ compared with from $Fr=0$ (the rigid lid model) to $Fr=0.2$. This nonlinear increase lies on the reason that the Froude number acts on the equation system as a form of $Fr^2$. However, comparing across these conditions, in spite of the difference between each other, the general trend is for an increase in maximum power coefficient from one-dimensional array to two-dimensional.

What needs to be stressed is that although not presented here, our computation results show that 
the increase in $C_{PGmax}$ is found not only for $B_L=0.25$ but also for arbitrary blockages. For arrays with higher blockages, this increase is relatively smaller, while for lower blockages arrays which are the practical scenario especially in deep water, the increase is apparent as Fig. 9 shows. 
Nevertheless, the increase in the power coefficient slightly sacrifice the efficiency of arrays as more power means more severe wake mixing. For one-dimensional array (corresponding to case 2), the efficiency for maximum power is 0.54, while for two-dimensional array (corresponding to case 1), the efficiency becomes 0.51 when power coefficient get the maximum value at $B_L \approx 0.6$. 
However, this decrease in efficiency is smaller compared with power increase. More importantly, for the same power coefficient, the two-dimensional array has higher efficiency than the one-dimensional. For example, if we let the power coefficient of two-dimensional array just get the maximum value of one-dimensional, 0.854, the efficiency will increase to 0.68, which is a larger improvement compared with 0.54 for one-dimension array. For other given power, this increase in efficiency is still effective. The reason lies on higher offset from maximum power coefficient than one dimensional array. As we have discussed above, offset from $C_{PGmax}$ increases the efficiency greatly. 
\subsection{Optimal arrangement for finite number of turbines}
\begin{figure}
  \centerline{\includegraphics[width=8cm]{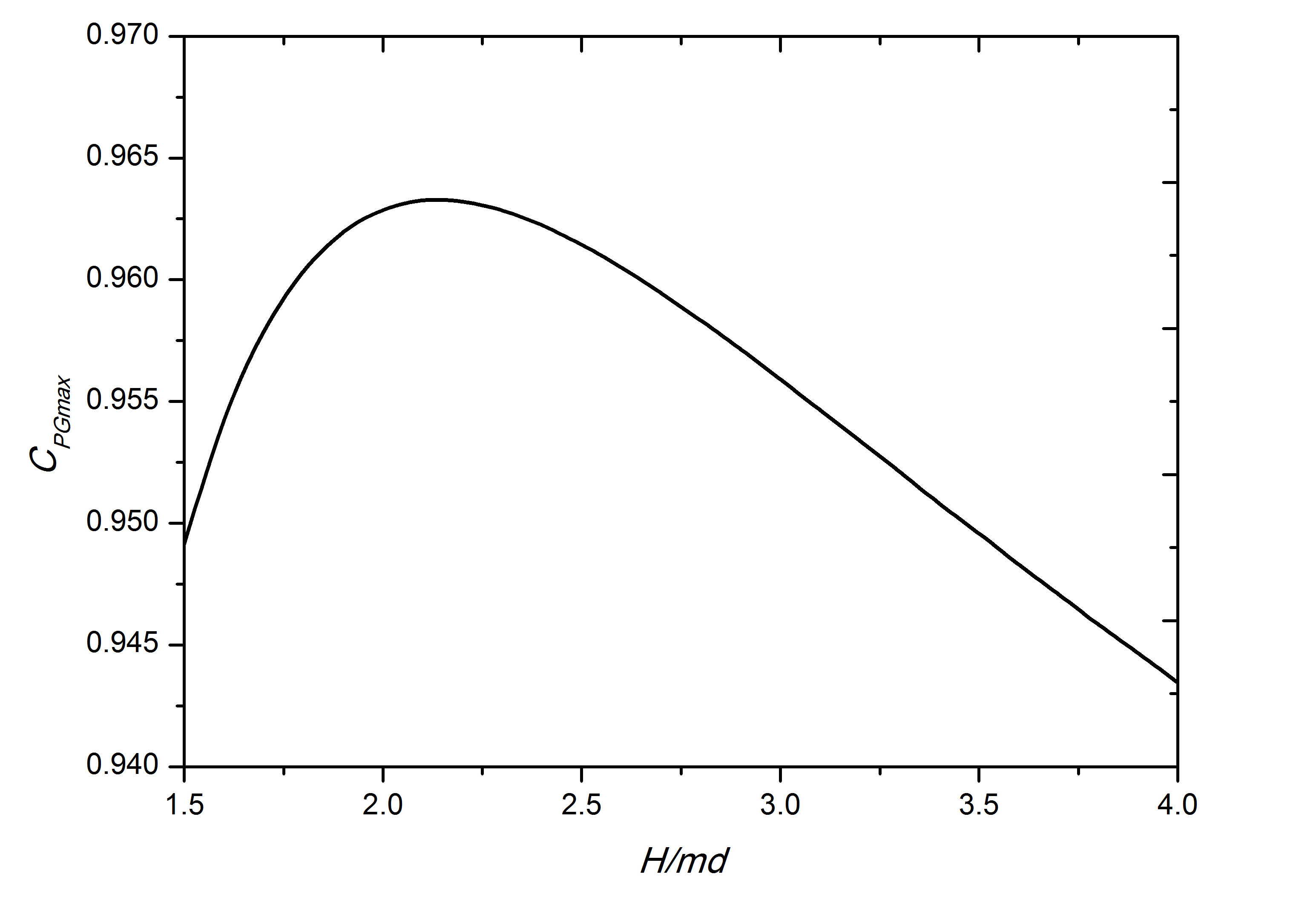}}
  \caption{Effects of $H/md$ and $W/nd$ on $C_{PGmax}$ for the fixed local blockage and global blockage. The results show the optimal condition of arranging finite number of turbines in both the lateral and vertical directions when intra-turbine spacing has been determined.}
\end{figure}

For two-dimensional array, an interesting topic here is that how to arrange finite number of turbines to acquire more power. Assumed that the installation method of the turbine array has been decided. The lateral intra-turbine spacing is fixed as half the diameter and the vertical is assigned as one fifth of the diameter considering the depth of channel. Thus the local blockage is approximately 0.45, a very likely value for small turbines. 
Now we consider the effects of  $H/md$ and $W/nd$ on the maximum power coefficient when the global blockage is fixed. For a channel, the fixed global blockage suggests that the total number of turbines with certain diameter is given, while  $H/md$ and $W/nd$ reflect how many turbines are arranged in vertical and lateral directions, respectively. 
If we plan to arrange 960 turbines of 5 m diameter in a channel of 3000 m width and 80 m depth, as Fig. 10 shows the present model predicts that the optimal  $H/md$ is approximately 2 ($Fr=0.2$). This means that the suggested arrangement is putting 8 turbines in vertical direction and 120 turbines in lateral direction from a power perspective. In this way thirty percent of width and sixty percent of depth are utilized. 
\subsection{arrays in infinitely wide channel}

\begin{figure}
  \centerline{\includegraphics[width=12cm]{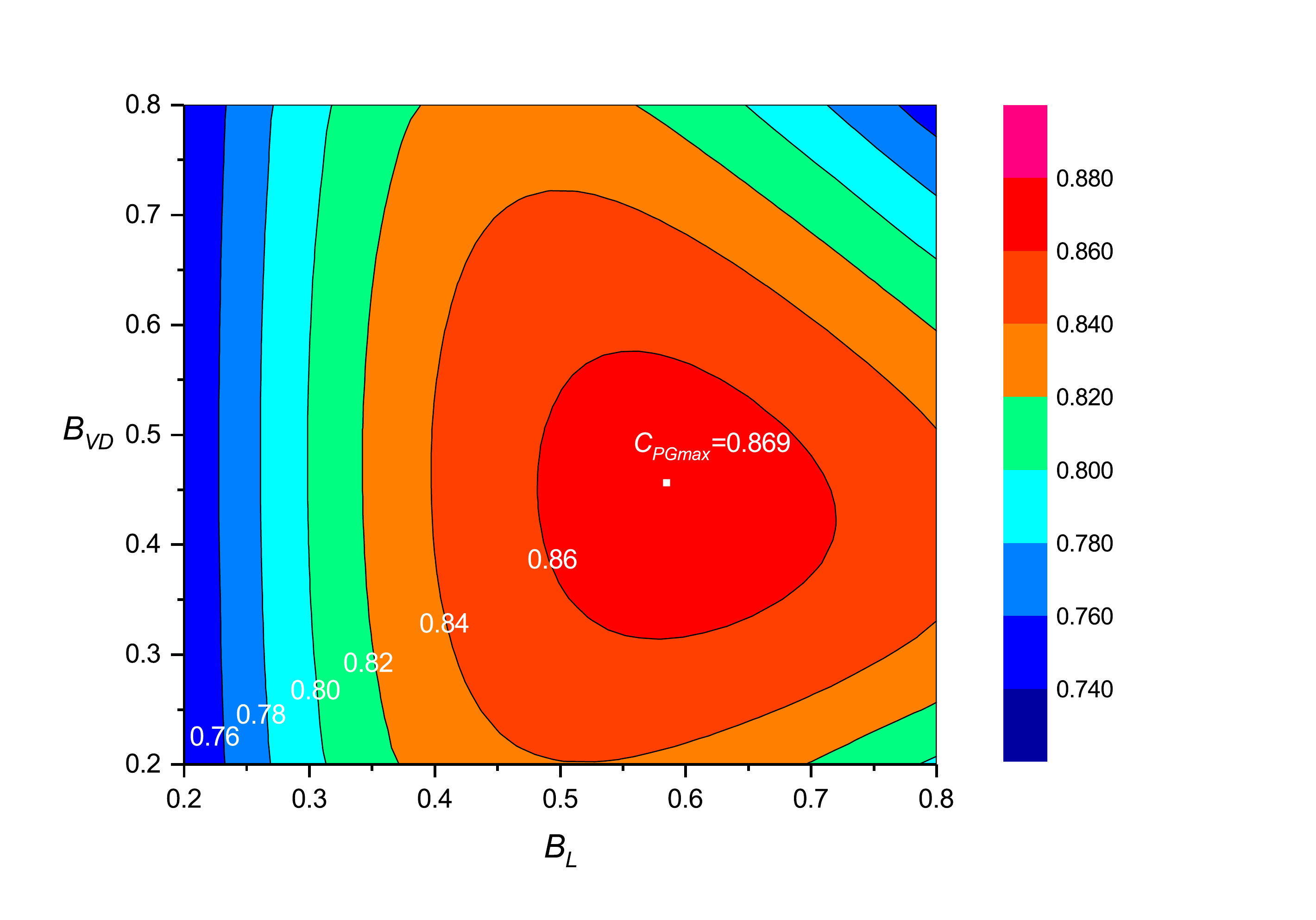}}
  \caption{The contour of $C_{PGmax}$ for arrays arranged on laterally unbounded flow. }
\end{figure}
Besides the performance of the arrangement described by finite parameters, we also concern the power of the array in an infinitely wide channel ($W/nd \to \infty$). This can be seen as the limiting case when array occupies an infinitesimal small region of a channel, which is practical for tidal arrays at present considering the restriction of  investment and geography. 
Also reflects the importance of this case is that the array in unbounded flow is hard to be analysed by experiments or numerical simulations. 
For an array in infinitely wide flow, we have $B_A = 0$ and $B_G = 0$, while
$B_L$ and $B_V$ are still finite values between 0 and 1.
Physically, the near array flow (including the local scale flow and vertical scale flow) are still affected by the arrangement of turbines while the flow far away from the array is no longer. Mathematically, for new equation system, compared with the model describe in Sec. II B, part (iii), the array scale problem are simplified as 
\begin{gather}
\xi_{2A}=1-\frac 12 Fr^2 (\alpha_{2A}^2-1),\\
\xi_{3A}=1-\frac 12 Fr^2 (\alpha_{3A}^2-\alpha_{4A}^2 ),\\
\alpha_{2A} \xi_{2A} =\alpha_{3A} \xi_{3A},\\
C_{TA}=2 \alpha_{2A}(1-\alpha_{4A}),
\end{gather}
together with the same Eq. (42).

Allowing both $B_L $ and $B_{VD}$ to vary, the contour of $C_{PGmax}$ (Fig. 11) demonstrates that there exists an optimal arrangement where $B_L \approx 0.6$ and $B_{VD} \approx 0.45$ to get an maximum power coefficient of 0.869 for $Fr=0.2$, which is a considerable increase compared with the results of two-scale rig lid model (NW12 model), $C_{PGmax}=0.798$ at $B_L \approx 0.4$ and more obviously with Lanchester–Betz limit of 0.593. This increase can be attribute to more complicated flow out of the turbine. Also, we calculate the maximum power coefficient for $Fr=0.3$, 0.874 and for rigid lid model, 0.865, suggesting that the free surface effects contribute very small for power increase compared with changing array configuration. This is because turbines are arranged in infinite flow which neglects the array scale depth change considering the tiny influence of array on whole flow (while the near turbine flow still has free surface change).  

\section{Numerical Simulation}
The two-scale model has been tested by numerical simulation and experiments. \citep{Nishino2013, Cooke2015} The separation of the turbine scale and array scale flow is obvious. 
Consider this together with the difficulty of calculating full array (To totally simulate an two-dimensional turbine array in a large channel, we need put at least hundred of turbines in the computational domain), now we only focus on the separation of vertical and lateral wake mixing of the array, which is an extremely important assumption in this model. What needs illustration is that for this numerical simulation, we do not try to figure out the performance of two-dimensional array with exact value, but to verify this basic assumption by modelling a plate in flow. 
\begin{figure}
  \centerline{\includegraphics[width=9cm]{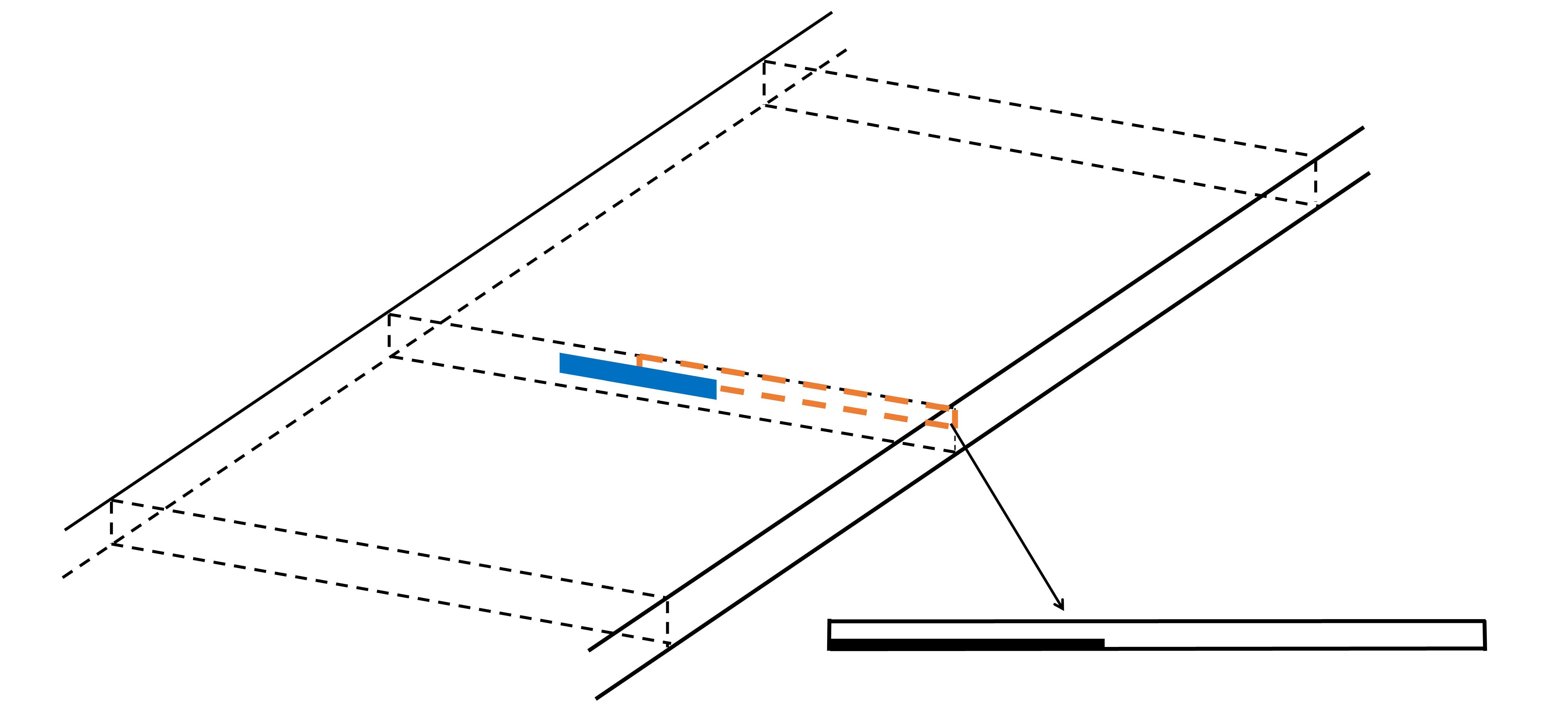}}
  \caption{Schematic of the computational examples. The computational domain is one quarter of the flow domain.}
\end{figure}
\subsection{Model description}
The numerical code was developed in Open FOAM 2.0. The governing equations are the incompressible RANS equations where the Reynolds Stress term is modeled using the standard k-epsilon model. The PIMPLE algorithm was used for pressure-velocity coupling. Corresponding to the theoretical model, the plate here is considered as a loss of momentum in the streamwise direction. The change in momentum equation to represent the plate is calculated as $Ku|u|/\Delta$, where $K$ is the strip resistance, $\Delta$ is the streamwise dimension of the plate.

Cartesian coordinates $(x, y, z)$ are employed, respectively representing the streamwise, vertical and lateral directions. The original of the coordinate is at the center of the inlet boundary. The computational case is a channel with twenty meters height and six hundred meters width. Thus the ratio of the channel cross-section is 30:1. The uniform streamwise velocity 2 m/s is set for the inlet boundary and a zero-gradient condition is applied at the outlet boundary. 

A plate of $l$ in lateral direction and $h$ in vertical direction is placed at $12l$ from upstream boundary in a channel having a streamwise length of $30l$. The plate center is just the center of the cross section (Fig. 12). The streamwise dimension $\Delta$ is 3 m and the changing of $\Delta$ around this value do not influence the final results. The streamwise length of mesh near plate is 0.25 m. Preliminary computations have shown that under such conditions for one scale model (for example an square or round plate putting on a channel with square cross-section or a plate with 20 m height putting in the computational channel which blocks the whole height), the power coefficients are nearly the same with theoretical values. 
As the flow is symmetry in both the lateral and vertical directions, computations are only performed on one quarter of the flow domain where the symmetry boundary condition are applied to all four longitudinal sides. 
\begin{figure}
  \centerline{\includegraphics[width=14cm]{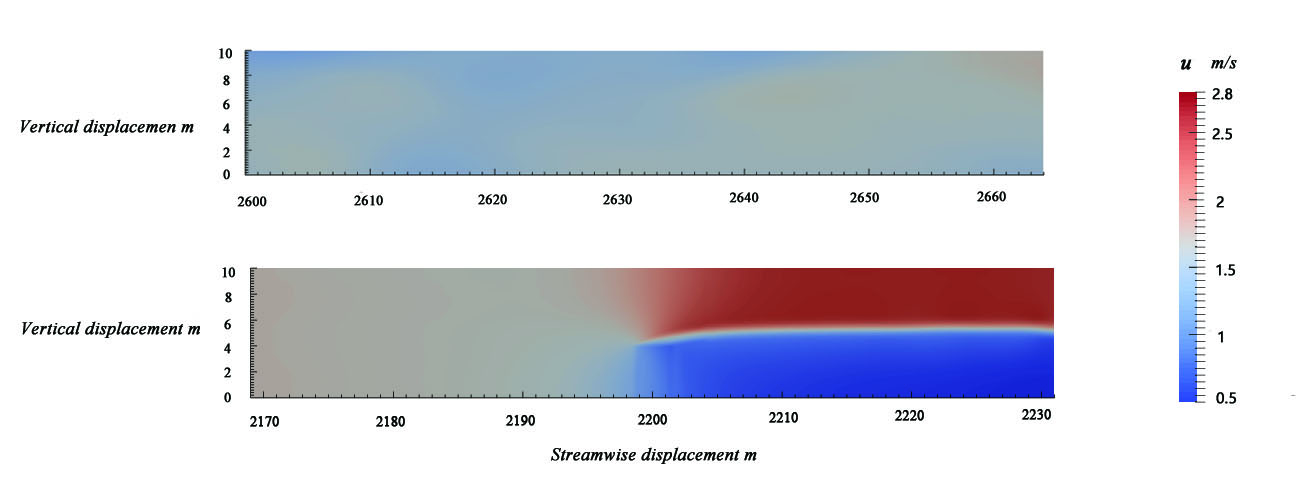}}
  \caption{Streamwise velocity on longitudinal plane ($y=50$ m)}
\end{figure}
\begin{figure}
  \centerline{\includegraphics[width=14cm]{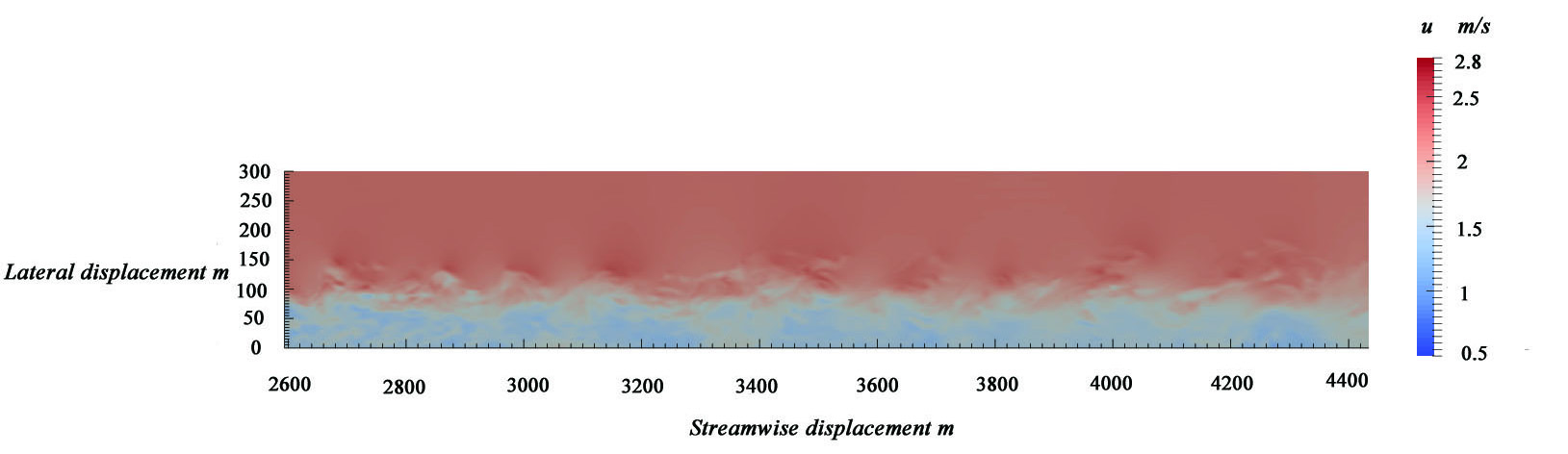}}
  \caption{Streamwise velocity on vertical plane ($z=5$ m) }
\end{figure}
\subsection{Results}

\begin{figure}
  \centerline{\includegraphics[width=9cm]{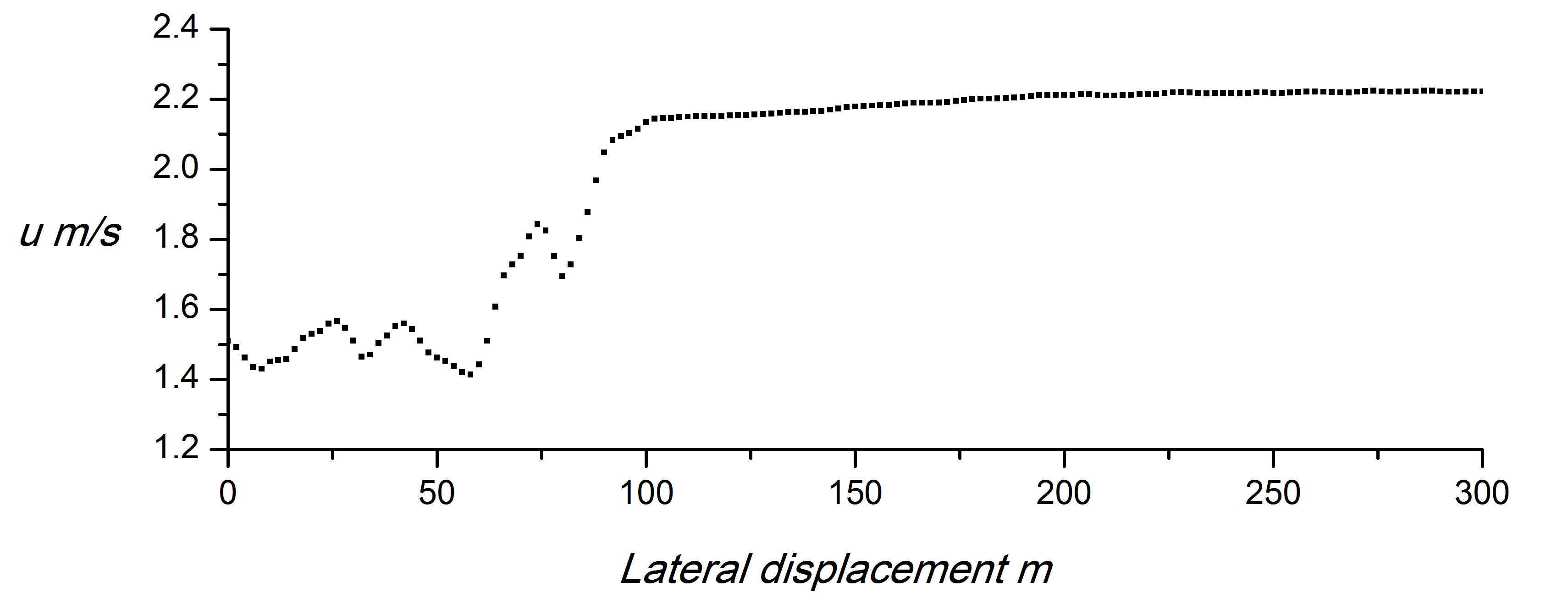}}
  \centerline{(a)}
  \centerline{\includegraphics[width=9cm]{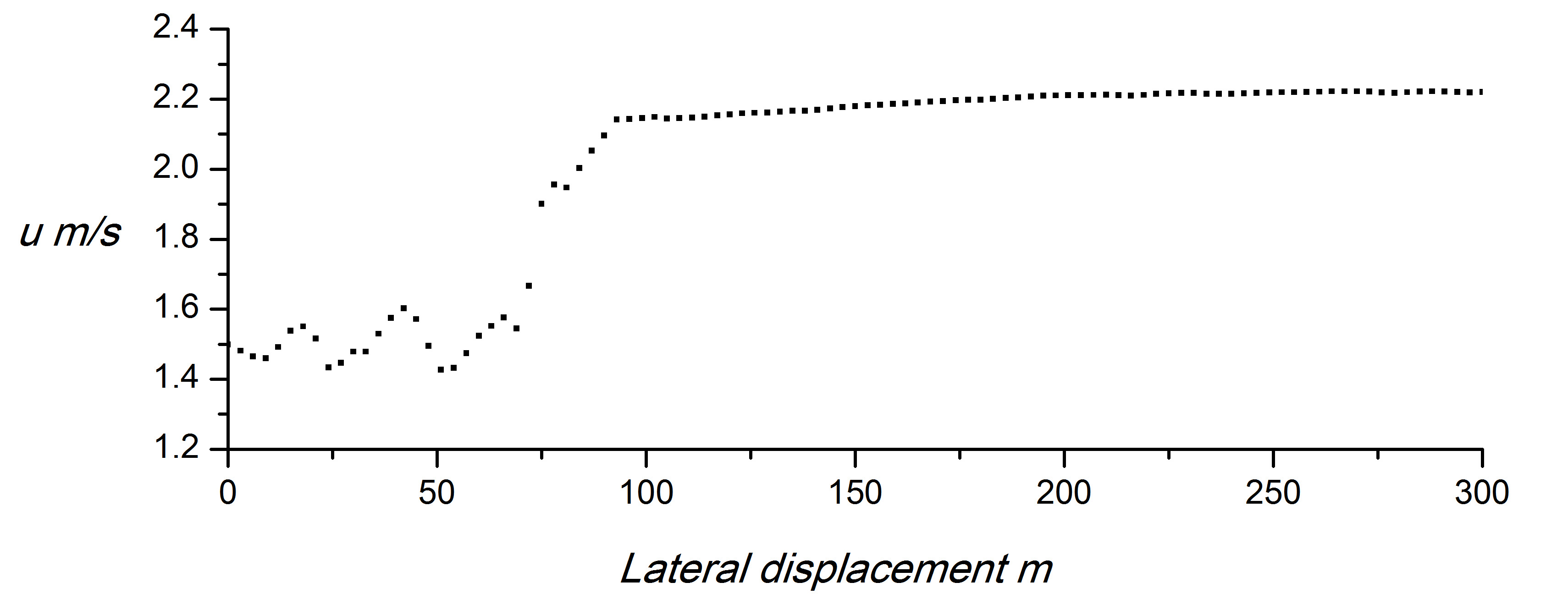}}
  \centerline{(b)}
   \centerline{\includegraphics[width=9cm]{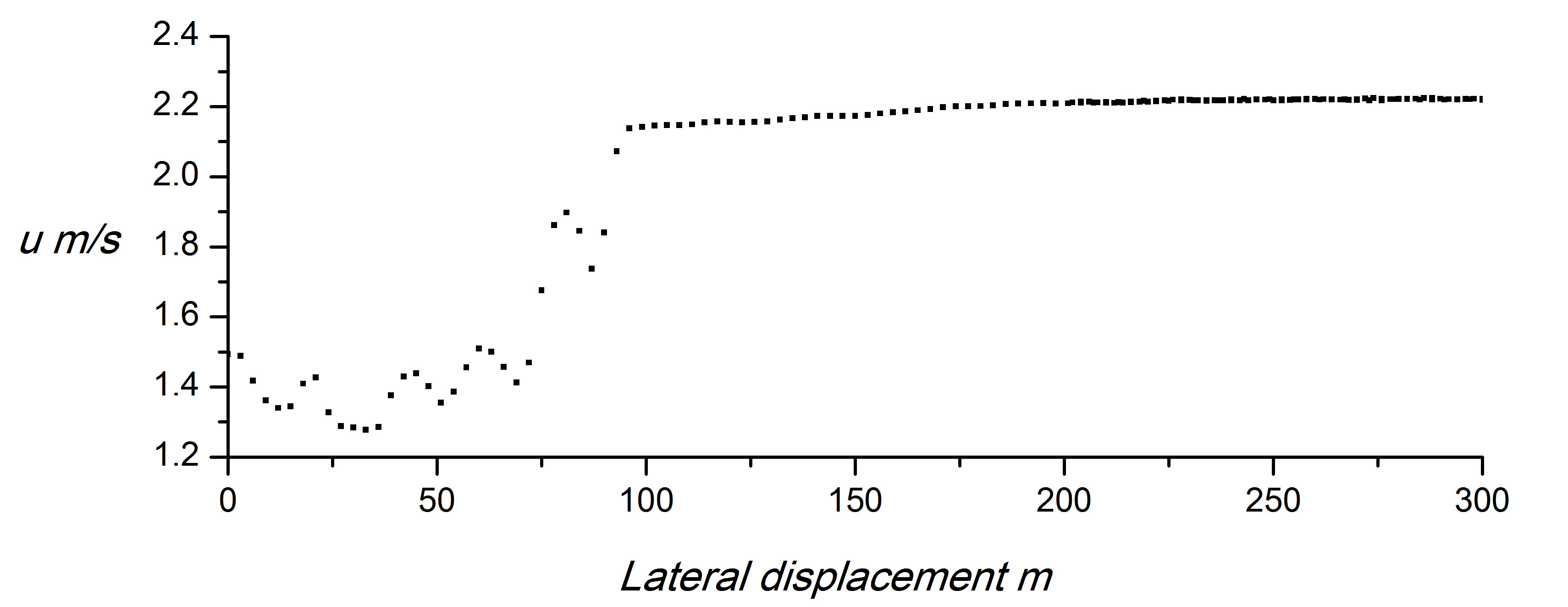}}
  \centerline{(c)}
  \caption{Streamwise velocity profile at $x=2620$ m. The heights are respectively (a): $z=2$ m, (b): $z=5$ m, (c): $z=8$ m.}
\end{figure}

Results are presented for a plate with $B_A=0.3$ ($l=180$ m) and $B_V=0.4$ ($h=8$ m) placed at $x=2200$ m. The separation of different scale is apparent and the vertical scale mixing is far more advanced than lateral scale ones. Fig. 13 show streamwise velocity contours on longitudinal plane ($y=50$ m) for an computational example with $K=6$. It is obvious that the vertical scale mixing is almost finished at 2600 m. However, the lateral scale separation continues to a far longer distance (Fig. 14). To further illustrate this, the streamwise velocity profiles at $x=2620$ m are shown in Fig. 15. For $y>90$ m, the velocity in all three height is approximately 2.2 m/s. For $y<90$ m, the velocity is uneven but mostly confined to $1.3-1.6$ m/s. This could be seen as a point which separates the major parts of vertical scale mixing with lateral scale mixing.

Also we can see that in this case, the vertical scale mixing is not totally completed before the starting of lateral scale mixing. The velocity is uneven because of remaining vortex. The quantitative results show that the increased power coefficient is approximately half of the expected increase according to theoretical results. 
In theoretical mode, we assume the height of the channel is negligible compared with its width. Under such conditions, it is reasonable to deduce that the vertical scale mixing is finished before the starting of the lateral scale mixing. In this computational case, the ratio of the channel is 30:1. It may be not enough for totally scale separation. For a natural deep channel, the ratio can be 100:1 thus the separation would be more apparent. 

\section{Channel Dynamics}

We have discussed the performance of the array under constant velocity. Yet the current in a channel would been influenced by the installation of turbines. The qualitative compare of the power of different arrangements can be obtained by comparing the efficiency or power coefficient under the same thrust (which means the same velocity). However, quantitatively, we must consider the dynamics of the array in a larger scale system.

Following GC05, the dynamics of the channel can be described by shallow water equations. The arranging of turbines is modeled as an added drag. As we concern the overall performance of the array, the integral form of the momentum equation would be used. Assuming a sinusoidal head loss between the two sides of the channel $ASin\omega t$  and ignoring the separation of the channel exit. The non-dimensional version of the governing equation of a channel with length $L$, width $W$ and height $H$ is: 

\begin{equation}
\dfrac {\partial u'}{\partial t'} = \sin t' - (\lambda_{D} + \lambda_{F}) |u'| u'
\end{equation}
where $u'=u / u_{max}$ is the dimensionless velocity divided by peak flow in frictionless channel. $\lambda_D=\alpha C_D L / H$ and $\lambda_F=\alpha C_F$  are non-dimensional drag coefficient where $\alpha=g A / \omega^2 L^2$. $C_D$ is the bottom drag coefficient and $C_F$  is the drag coefficient caused by turbines. The time $t'$ is non-dimensionalized by $1/\omega$ therefore measured in radians. It should be noted that for simplicity, here we set a constant cross section and constant turning.

Assume the covered area of turbines and their wakes are smaller compared with the whole channel area, thus the bottom drag is not influenced by the arrangements of turbines. According to the definition of drag coefficient, the turbine drag coefficient $C_T$ can be determined by actuator disc model:
\begin{equation}
C_F= \frac{1}{2} N_R B_A C_{TA}
\end{equation}
where $N_R$ is the number of rows.

The average non-dimensional power is calculated as:
\begin{equation}
\overline{P'}=\frac{\int_t^{\frac{2 \pi}{\omega}+t} P \,\mathrm{d} t}{\int_t^{\frac{2 \pi}{\omega}+t} \frac{1}{2} \rho |u_0|^3 A_T \,\mathrm{d} t}=\frac{C_{PG} \overline{|u'|^3}}{\overline{|{u_0}'|^3}}
\end{equation}
where $u_0$ is the undisturbed velocity before arrangement of turbines, $A_T$ is the total area of turbines. Here the non-dimensional power represents the ratio of the power per turbine to the kinetic flux of the flow in the same area before arraying any turbines in the channel.

\begin{table}
\caption{Parameters of the three hypothetical channels}
\begin{ruledtabular}
\begin{tabular}{lccccc}
        & Depth, H   & Length, L  & Width, W & $\alpha$ &  $\lambda_D$        \\[10pt] \hline
       Tidal Strait   & 100 m & 50 km  & 10 km & 0.3 & 0.35 \\
       Medium Channel   & 50 m & 20 km  & 9 km & 1.1 & 1.1 \\
\end{tabular}
\end{ruledtabular}
\end{table}

Under this larger scale system, we re-compare examples of one-dimensional array and two-dimensional array in Sec. III B. The background channel is the tidal strait in \citet{Vennell2010} and the medium channel in \citet{Vennell2016}. The parameters of the channel is shown in Table I. Corresponding to the above analysis, we still assume that one row of turbines are put in the channel. As we neglect the free surface change in channel equations, the rigid-lid model is used for array problem.

\begin{figure}
  \centerline{\includegraphics[width=11cm]{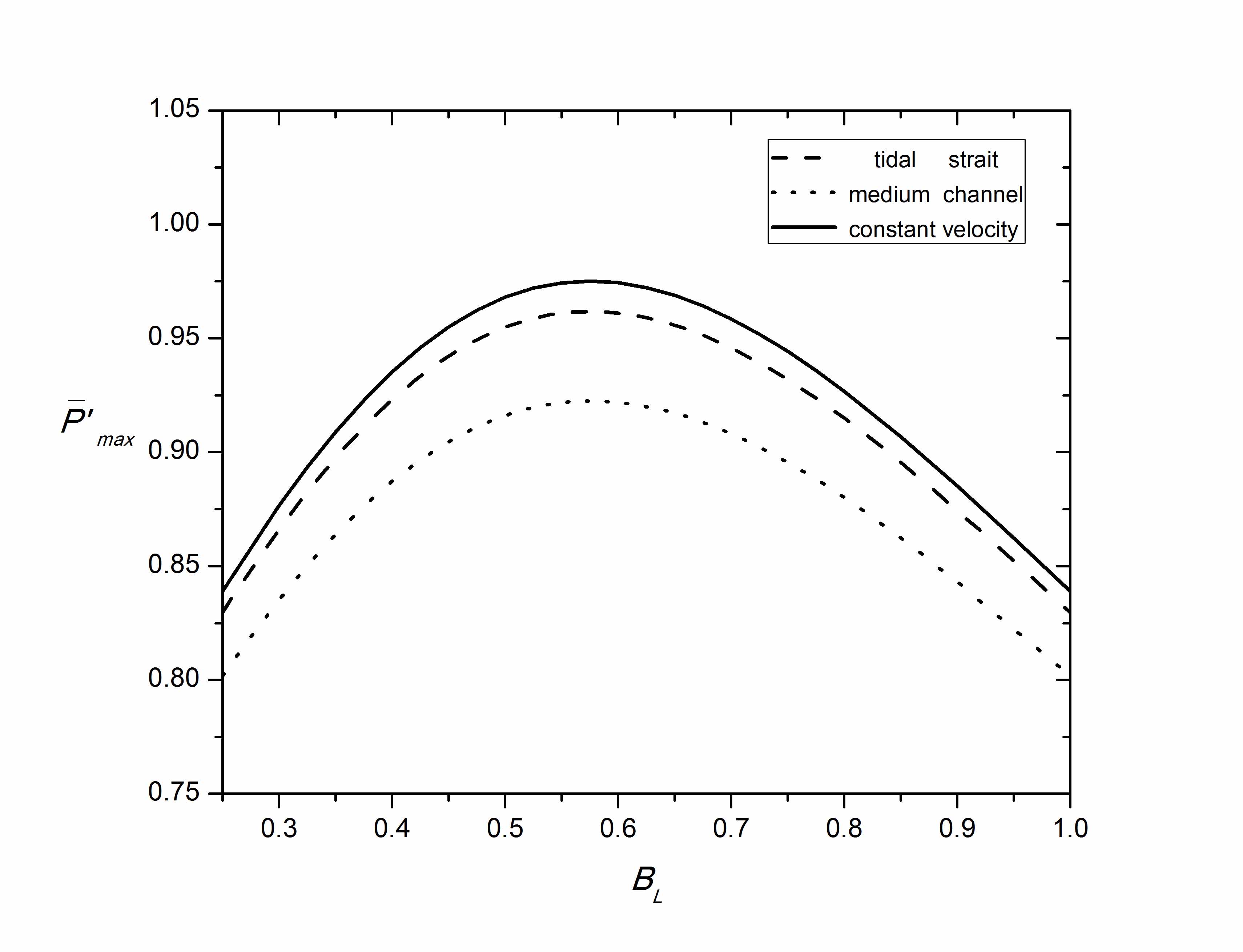}}
  \caption{Variation of maximum non-dimensional power of array with local blockage for $B_G=0.075$ and $B_A=0.3$ considering the channel dynamics}
\end{figure}

The maximum values of the non-dimensional power for two-dimensional array in hypothesized tidal strait and medium channel are shown in Fig. 16. Also presented here are the results which do not consider the change of velocity. The constant velocity value is actually the $C_{PG}$ in Sec. III. When consider the channel dynamics, the predicated power is lower than of constant velocity. For the tidal strait (or large channel), the decreased value is tiny as the turbine force is small compared with the inertia of the channel, while for the medium channel, the decrease value is noticeable. Yet the general trend is still that the predicated power of two-dimensional array is larger than one-dimensional and optimal situation is unchangeable. 

Besides the absolute power, the power increase from one-dimensional array to optimal two-dimensional array ($B_L=0.58$) also slightly diminishes when consider channel dynamics. The predictable power increases $16.2\%$ from one-dimensional array situation ($B_L = 0.25$ and $B_L = 1$) to optimal two-dimensional array situation when velocity is constant. For the array in tidal strait and medium channel, this increase is $15.9\%$ and $15\%$ respectively, which is predictable as the power increase is at the expense of force increase. The increased power will be partly offset as the increased force cause lower velocity and this effect is more obvious for channel with relatively smaller inertia. However, for situations we have discussed, this offset is limited. We can still benefit a lot from putting turbines in two-dimensional array. For arrays with several rows in small channel, the offset will be more noticeable and need attentions.  

\section{Discussion}

The original two scale momentum theory has been extended to include the free surface effects and allow arraying turbines in vertical direction. Considering the strong constriction of the flow between surface and channel bed, an additional scale created by separation of vertical and lateral wake mixing is built in present work. The numerical simulations demonstrate that the separation of wake mixing in these two directions is apparent.
Solutions for the power coefficient, efficiency and depth change of two-dimensional array have been obtained. 
The results show that the new arrangement can effectively increase the power coefficient and efficiency compared with one-dimensional array. Furthermore, the influence of  $H/md$ and $W/nd$ on maximum power coefficient has been analysed to figure out how to arrange finite number of turbines in two directions. Also shown here are the results of array in infinite wide channel. When $Fr=0.2$, the maximum power coefficient increase from 0.798 for one-dimensional array in NW12 model to 0.869 in present work. When we consider the reduced velocity, the power produced by turbines and the increased ratio would be smaller than predicated values under constant velocity. But the benefit from arranging turbines in two-dimensional array is still noticeable and the decreased value is small for large or medium channel with finite turbines.

In modelling process, herein the boundary of the local scale flow passage  have been assumed to be constant, provided that : (i) the change of the height is neglected in the local scale domain, and (ii) the number of turbines is sufficient large. In practical, the cross-section of local flow passage could not be completely rectangular, partly because of flow expansion when the number of turbine is not sufficient large, which is possible especially in vertical direction consider the limitation of depth. Qualitatively the relatively short array show the same trend with the long one and quantitatively the difference is smaller than ten percent \citep{Nishino2013}, so the results are still reliable as we concern trend of parameters more than values. 
Besides flow expansion, gravity relating with the water depth change is also an important factor. The height of local scale flow is not strictly same along flow direction. This may be analysed by consider the different force between each local scale flow passage, however greatly adds the complexity of present model. The results show that the free surface change is small compared with the water depth. Thus the change in each local scale passage is further smaller. So, neglecting the added force in momentum equation is also a reasonable approximation. 
Also note that the basic assumption for these three scale separations is that the clearance between free surface and channel bed is far smaller than width of channel. Actually, for most of channels this assumption is valid, while for channels which width is only several times larger than height, the flow may become two scales, just local scale and array scale. As numerical simulations show that two-scale flow is existing, it is reasonable to assume that the three-scale separation would occur in new model under similar assumptions. Of course, further numerical or experimental work is needed to validate the three-scale flow separation.

Moving to the solving method, different with traditional one by choosing local scale induction factor as the known variable, this study specifies vertical or array scale thrust coefficient to get the value of larger scale problem then smaller. We analyse the results and chose values of zone I which satisfy subcritical requirement and slowing down requirement as desired solutions. Actually, it is radical to draw a conclusion that theoretical model can get a supercritical solution as \citet{Whelan2009} and \citet{Houlsby2008} did. In momentum model we assume the conservation of energy in bypass flow. In contrast for this assumption, practically the jump between subcritical flow and supercritical flow would cause great energy loss. So, although the subcritical solution has been verified by experiments \citep{Whelan2009}, the physical interpretation of the free surface model still need to be investigated further.

The results of this work together with existing literatures show that for fixed global blockage (or same total area of turbines),  the arrangements ranking based on the maximum power coefficient is: two-dimensional array (investigated by three-scale model) $>$ one dimensional array (investigated by two-scale model) $>$ one single turbine. This implies a possible trend that turbines would have higher potential power when they arranged as more complicated array in more complicated flow environment (flow separation). Of course, stricter assumptions are needed if we continue to enhance the power coefficient in this way. However, these do not weaken the meaningful of this trend.

Finally, we notice that the bed friction is neglected within array scale dynamics. For large array, the changing of friction drag may be important and  the quantitative influence of friction on turbine performance has been investigated by \citet{Creed2017}, showing an increase in power as friction enhances. The work of \citet{Creed2017} assumes constant velocity in vertical direction. Actually in smaller scale problem, the real flow is shear vertically because of the presence of bed friction. This may led to change of effective blockage for vertical scale problem as discussed in \citet{Draper2016}. The bed friction can also influence the channel dynamics considering force change within the array. Neglecting the inertial effects, \citet{Gupta2017} found that the bed friction would reduce the velocity and more comprehensive work is needed to analyse it in the future.

The authors would like to acknowledge the support of the State Key Laboratory of Ocean Engineering, University of Shanghai Jiao Tong University and the Thousand Talents Plan.

\nocite{*}
\bibliography{aipsamp}
\end{document}